# Exploring the redundancy of Radon transform using a set of partial derivative equations: Could we precisely reconstruct the image from a sparse-view projection without any image prior?

**Xuanqin Mou, Jiayu Duan**

Institute of Image Processing, School of Information and Communications Engineering
Xi'an Jiaotong University, Xi’an, China

E-mail: xqmou@mail.xjtu.edu.cn

**Abstract**

In this study, we propose a universal n-th order partial differential equation (PDE) of 2-D Radon transform to disclose the relationship of Radon transform over a neighbourhood of integral line. Specifically, a CT geometry of dual centers of rotation is introduced to formulate an object independent PDE that presents the local correlation of Radon transform on the variables of distance and angle, named LCE (local correlation equation). The LCE is directly available to divergent beam CT geometries, e.g., fan beam CT or cone beam CT. In this case, one rotation center can be set at the focal spot, so that the LCE becomes a general PDE for the actual CT systems with single rotation center (origin). Thus, we deduce two equivalent LCE forms for two widely used CT geometries, i.e., cLCE for circular trajectory and sLCE for stationary linear array scanning trajectory, respectively. The LCE also explores the redundancy property existed in Radon transform. One usage of the LCE is that it supports a sparse-view projection could contain enough information of complete projection, and hence projection completeness in CT scanning would be no longer needed. In this regard, based on the circular scanning trajectory, we explore whether the LCE is able to solve sparse-view problem without the help of image prior by proposing a discrete LCE based interpolation method. Moreover, a unified reconstruction framework combining a regularized iterative reconstruction with the LCE based interpolation is also proposed to cope with higher sparsity level and the discrete LCE solving process. In experimental validation, we chose 1/4 and 1/8 sparsity to verify the discrete LCE interpolation method and the unified reconstruction scheme, respectively. The results confirm that a sparse-view projection is able to contain enough information of complete projection of CT. Sparse-view CT to a certain extent is feasible to realize a perfect reconstruction as from complete projection. It would be expected that combining the LCE property will boost the research on CT theory and development of CT geometry design in the future.

## 1. Introduction

Computed tomographic (CT) involves in Radon transform and its inversion computation. Mathematically, the Radon transform maps a function into a set of its line integral over a hyperplane, while in practice the function is generally a representation of the inner structure of an imaging object. With the inversion of the Radon transform, the inner structure of the imaging object can be revealed. There are numerous ways to calculate the inversion of the Radon transform. For instance, based on the property of the Fourier transform, Ronald derived the central slice theorem (Bracewell, 1956) that is the core of the filtered back projection (FBP) algorithm (Bracewell and Riddle, 1967) in 2-D CT and Feldkamp-Davis-Kress (FDK) algorithm (Feldkamp *et al.*, 1984) in 3-D CT. These algorithms are categorized as analytical algorithm. CT can also be regarded as an inverse problem, which can be optimized iteratively based on multiple strategies. In CT, Algebraic Reconstruction Technique (ART) and its variants are of the most common iterative reconstruction technique (Gordon *et al.*, 1970; Yu and Wang, 2012).

In CT, the precise of the reconstruction is paramount. The precision relates to the acquisition number and trajectory. Higher sampling numbers in CT scanning generally result in better image quality and higher spatial resolution. With sufficient sampling number, trajectory design further ensures the data completeness. Specifically, for 2-D image reconstructions, it is generally accepted that accurate and stable reconstruction of any region of interest (ROI) requires that all projection of the integral lines passing through the object are known (Kak and Slaney, 2001; Noo *et al.*, 2002). In 3-D reconstruction, the projection should satisfy the Tuy-Smith condition (Tuy, 1983; Smith, 1985) to guarantee the precision of the reconstructed image, which states that every plane through the ROI must intersect the scan trajectory. When the completeness condition is violated, there leads to the problem of sparse-view CT (less sampling number) or limited-angle CT (scan trajectory is truncated).

In practice, it is challenging to ensure the data completeness because of the variant constraints. For example, due to the size and weight of the traditional X-ray source, it is impossible to generate rapidly scan. To pursuit temporal resolution, cutting down the sampling number is inevitable. Besides, the balance between the spot size and data completeness is also tricky. The smaller spot size means higher spatial resolution but longer exposure time, which conflicts with the high temporal requirement in medical X-ray CT diagnosis.

Moreover, in some circumstances, it is necessary to choose incomplete projection during scanning. In medical diagnosis, temporal resolution and the ALARA principle are crucial in examination. Hence, in order to minimize the radiation dose and increase the temporal resolution for ALARA consideration, researchers proposed sparse-view scanning protocol (Sidky *et al.*, 2006; Xu *et al.*, 2012; Kudo *et al.*, 2013) Meanwhile, the scanning geometry or the choice of X-ray source may also generate sparse-view problem. For example, in stationary tomosynthesis, the CNT-based X-ray tube suffers from the unneglectable distance between individual CNT cells, causing the sparse-view problem during measurement (Inscoe *et al.*, 2022). In industrial CT, the rod or irregular objects limit the scanning trajectory, which causes sparse-view problem (Fu *et al.*, 2007; Novak *et al.*, 2020) .

In general, images reconstructed from sparse-view problem face the severe streak artefact, which degrades the image quality (Lu *et al.*, 2011; Xu *et al.*, 2012; Bao *et al.*, 2019). At present, the most efficient way of solving the sparse-view reconstruction is to introduce image priors by compressive sensing (CS) or deep learning techniques. The idea of CS is that natural images, including medical X-ray anatomic images, contain redundancy which makes possibly reconstruction of the image from a down-sampling data, while CS helps to realize the sparse-view reconstruction computation. By CS, sparse-view CT reconstructions can achieve satisfied results in some cases, or at least the artefacts will be suppressed (Kudo *et al.*, 2013). Nowadays, commonly used CS technics include total variation (TV) (Sidky *et al.*, 2006; Liu *et al.*, 2012; Bian *et al.*, 2013; Niu *et al.*, 2014), dictionary learning (Xu *et al.*, 2012; Bao *et al.*, 2019), tensor representation (Zhang *et al.*, 2016; Hu *et al.*, 2020), etc. Deep learning networks known as multi-layer feature representation schemes, having higher efficient information extraction ability, have also been applied for sparse-view CT reconstructions (Chen *et al.*, 2018; Chen *et al.*, 2020; Su *et al.*, 2021; Zhang *et al.*, 2021).

Another way to deal with the sparse-view CT problem is interpolation of projection data. Projection data, also known as sinogram of CT, having redundancy is a common sense (Medoff *et al.*, 1983). Indeed, the invention of iterative reconstruction for CT was to deal with the reconstruction of incomplete projection data since iterative reconstruction can estimate unmeasured views by reprojection process, to some extent the redundancy in sinogram was used (Medoff *et al.*, 1983). However, the redundancy in CT sinogram is far from being thoroughly utilized in CT reconstructions including the sparse-view CT problems.

In CT sinogram, projection data is defined with regard to variables of view angle and signed distance of the projection plane from the origin. Specifically in 2-D sinogram, the variables are the view angle and the distance of the projection line from the origin (see Fig. 1 in IIA). Hence, sinogram interpolation means to recover an unmeasured projection from the measured projections of its neighbouring variables, thus the sparse-view problem will be resolved. In this context, a mathematic model to explore the relationship of Radon transform over neighbouring sinogram will be necessary to the interpolation problem. Unfortunately, in present CT theory, this relationship is depended on the imaging object (Natterer, 2001) (see IIA for details), which is not universal and hence cannot be used for interpolation computation.

One attractive work of exploring the redundancy of Radon transform and implementing sinogram interpolation is John's equation based scheme for cone beam CT (CBCT) (Patch, 2002; Defrise *et al.*, 2003; Yan *et al.*, 2010). The named John's equation is a set of ultrahyperbolic partial differential equations (PDE) for ranged line integral of the imaging object, which disclose the relationship over the neighbouring ranged line integral. Hence, these equations can be used for estimating unmeasured views. However, the John's equation works under a specific trajectory condition based on which the recovery accuracy of central detector is much higher than the detectors in edge area (Patch, 2002; Defrise *et al.*, 2003; Yan *et al.*, 2010). Moreover, this method is specifically for CBCT. It is hard to employ the John's equation in other CT geometries. Above limitations hinder the John's equation from a universal solution to sparse-view CT problems.

Resorting to deep neural networks to handle with restoring sparse-view sinogram is recently a hot topic (Lee *et al.*, 2018; Ding *et al.*, 2021). Although deep learning technics can be expected, the application to recover sparse sinogram is difficult. Apart from common concerns of data dependency and generalization ability, sinogram is of a dimensionality reducing space and does not have structured characteristics, to which deep neural networks is not easy to represent and hardly used for interpolation.

Anyway, building a universal model to disclose correlation property of Radon transform over a neighbourhood of integral line is attractive. In this regard, we follow the differential constraint condition (DCC) (Natterer, 2001; Tang *et al.*, 2011) to derive a universal PDE of Radon transform. Specifically, to tackle the limitation of DCC that is image object dependent, we propose a CT geometry of dual centers of rotation and derive an any order PDE that Radon transform meets. The derived PDE is called as local correlation equation of Radon transform (LCE). The LCE is image object independent and hence reveals a universal property of Radon transform. Then we give the equivalent LCE forms of two popularly used CT geometry, i.e., cLCE for circular scanning trajectory and sLCE for stationary linear array scanning trajectory. To show the feasibility of using the proposed PDE to solve sparse-view problem, we propose a sinogram interpolation scheme for circular scanning CT geometry by using the proposed discrete first order cLCE. Experiments in this study show that the proposed interpolation method can precisely recover sinogram from 1/4 sparse-view projection. We also propose a unified iterative reconstruction framework by combining an iterative reconstruction method with the proposed interpolation scheme. The unified method achieves satisfied reconstruction result from 1/8 sparse-view projection.

The contribution of this study lies in threefold. 1) To our best knowledge, the proposed PDE is the first universal model to reveal local correlation property of Radon transform. With the universal property of the PDE, the redundancy existed in Radon transform can be freely investigated. It would benefit to further development of CT theory. 2) The redundancy of Radon transform can be utilized by the proposed PDE to recover the sinogram from sparse-view scanning, which supports that in theory projection data completeness in terms of enough scanning number is no longer necessity. 3) The integrated iterative scheme proposed in the study suggests that future CT reconstruction framework should incorporate the PDE to handle many challenges in CT by extensively exploring the redundancy of Radon transform. The future work is not only for sparse-view CT and low-dose CT, but also for benefiting to construct interpretable deep neural network to process various CT problems, and etc.

The remainder of this paper is organized as follows. Section II describes the mathematical deduction of the PDEs of Radon transform. Meanwhile, a validation of the proposed PDEs is also provided based on numeric simulations. In Section III, we propose a sparse-view interpolation scheme for circular CT trajectory, as well as unified iterative reconstruction scheme by combining the proposed interpolation scheme. In Section IV, we demonstrate the experimental results of the proposed sparse-view interpolation scheme and the unified iterative reconstruction method. In Section V, we discuss the related issues of this study and make conclusion.

## 2. Methods

### *A. Preliminary*

Recall the 2-D Radon transform shown in Fig. 1, Let $L(t,\boldsymbol{\theta})$ denote the line from $\xi$ to $\eta$ in the Euclidean plane with the origin $\mathbf{O}=(0,0)$ of the world coordinate, $\boldsymbol{\theta}=(\theta_1,\theta_2)=(\cos\theta,\sin\theta)$ is the angle of the normal vector to $L$, makes with the $x_1$ axis, and $t$ is the signed distance of $L$ from the origin. In a real CT device, the distance is normally discretized as a detector index.

$$
\begin{aligned}
L(t,\boldsymbol{\theta}) &= \{(x_1,x_2)\in\mathbb{R}^2 : (x_1,x_2)\cdot(\theta_1,\theta_2)=t\} \\
&= \{(x_1,x_2)\in\mathbb{R}^2 : (x_1,x_2)\cdot(\cos\theta,\sin\theta)=t\}
\end{aligned}
. \quad (1)
$$

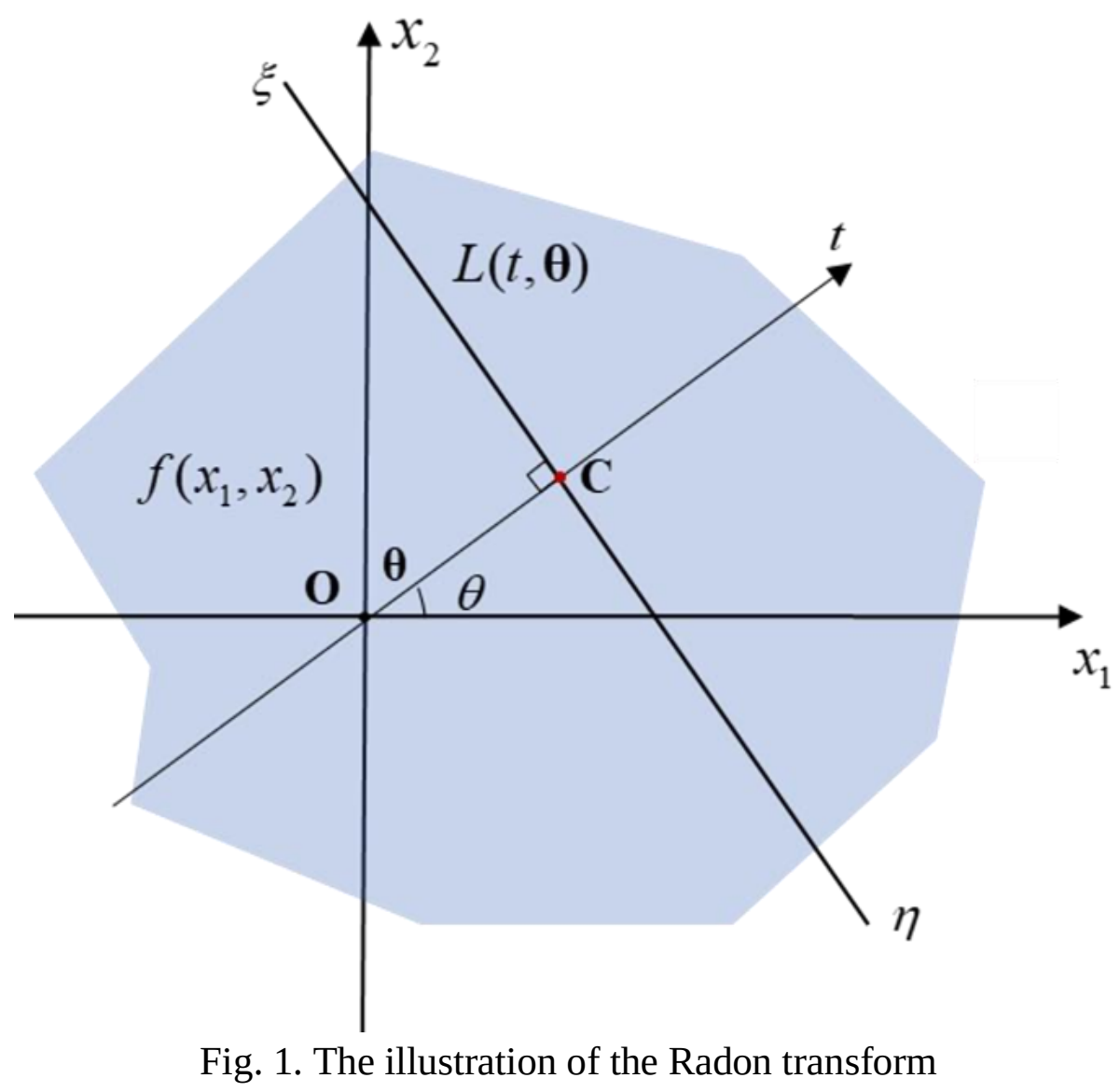

Fig. 1. The illustration of the Radon transform

The Radon transform along the line $L(t,\boldsymbol{\theta})$ can be described as:

$$
\begin{aligned}
R(\boldsymbol{\theta},t) &= \int_{-\infty}^{+\infty}\int_{-\infty}^{+\infty} f(x_1,x_2)\delta(x_1\theta_1 + x_2\theta_2 - t)dx_1dx_2 \\
&= \int_{-\infty}^{+\infty}\int_{-\infty}^{+\infty} f(x_1,x_2)\delta(x_1\cos\theta + x_2\sin\theta - t)dx_1dx_2
\end{aligned}
\quad . \tag{2}
$$

With $\delta(.)$ function defined as (Natterer, 2001):

$$
\delta^{b}(t) = \frac{1}{2\pi}\int_{-b}^{b} e^{its}ds\ , \tag{3}
$$

the relationship of partial derivatives of the projection between axis angle and distance can be denoted as (Natterer, 2001):

$$
\frac{\partial}{\partial\theta_k}R(\boldsymbol{\theta},t) = -\frac{\partial}{\partial t}(R(x_k f))(\boldsymbol{\theta},t), k = 1,2 \tag{4}
$$

In any direction of $\theta$, Eq.(4) can be further deduced as (Tang *et al.*, 2011):

$$
\begin{aligned}
\frac{\partial}{\partial\theta}R(\theta,t) &= \frac{\partial}{\partial\theta_1}R(\theta,t)\frac{\partial\theta_1}{\partial\theta} + \frac{\partial}{\partial\theta_2}R(\theta,t)\frac{\partial\theta_2}{\partial\theta} \\
&= \frac{\partial}{\partial t}(R(x\cdot\boldsymbol{\theta}^{\perp})f))(\boldsymbol{\theta},t)
\end{aligned}
\quad , \tag{5}
$$

where $\boldsymbol{\theta}^{\perp} = (\sin\theta, -\cos\theta)$ is the unit normal vector to $\boldsymbol{\theta}$, $\boldsymbol{\theta}^{\perp}\cdot\boldsymbol{\theta} = 0$. From Eq.(5), it is apparent that there is an implicit relationship between angle and distance differentials since the calculation of the right size of the equation needs knowing the object function $f$, which is object depended and hence hard to solve. In the next subsection, we introduced a geometry of dual centers of rotation to decouple the object from Eq.(5).

### *B. Object-free PDE deduction based on the geometry of dual centers of rotation*

To obtain a direct relationship between derivatives of the Radon variables, we propose a special CT geometry of dual centers of rotation, which is shown in Fig. 2.

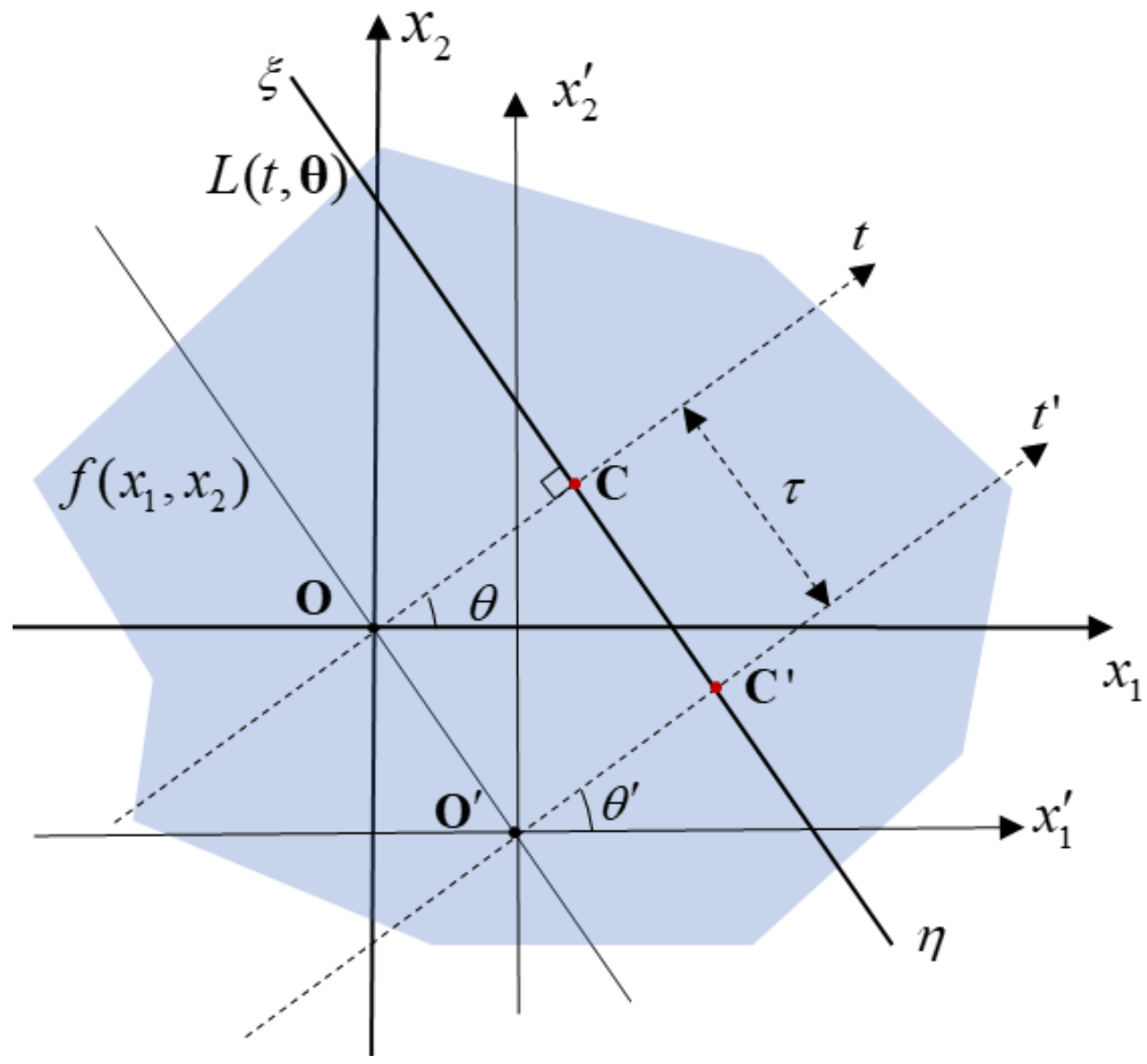

Fig. 2. The pseudo rotation centers illustration

The Randon transform along the line from $\xi$ to $\eta$ is defined by (2) regarding to the coordinate system $(x_1, x_2)$ with the origin $\mathbf{O}$ of the world coordinate. However, we can also define the same Radon transform regarding to a translational coordinate system $(x_1', x_2')$ with the origin $\mathbf{O}'$, while $\mathbf{O}' = \mathbf{O} + \tau\boldsymbol{\theta}^{\perp}$. Clearly, the angular derivative of the Radon transform with respect to $\theta$ is equivalent to the angular derivative of the Radon transform along $L(t,\boldsymbol{\theta})$ rotating at point $\mathbf{C}$. Equally, the angular derivative respect to $\theta'$ is equivalent to that at $\mathbf{C}'$. In this way, we call $\mathbf{C}$ and $\mathbf{C}'$ as the dual centers of rotation for the Radon transform along the line from $\xi$ to $\eta$. Let $R(\theta,t)$ denote the Radon transform at the origin of $\mathbf{O}$, and $R(\theta',t')$ denote that of $\mathbf{O}'$. We can observe that $R(\theta,t)$ and $R(\theta',t')$ represent the Radon transform of the same line integral when $\theta = \theta'$ and $t = t'$. Based on Eq.(5), we have:

$$
\begin{aligned}
\frac{\partial}{\partial\theta}Rf(\theta,t) &= \frac{\partial}{\partial t}(R(x\cdot\boldsymbol{\theta}^{\perp})f))(\theta,t) \\
\frac{\partial}{\partial\theta'}Rf(\theta,t') &= \frac{\partial}{\partial t'}(R(x\cdot\boldsymbol{\theta}^{\perp})f))(\theta,t')
\end{aligned}
\quad , \tag{6}
$$

where $\frac{\partial}{\partial\theta}Rf(\theta,t), \frac{\partial}{\partial\theta}Rf(\theta,t')$ represents the angular partial derivative of the Radon transform along the rotation centers C, C', respectively. Because $R(\theta,t)$ and $R(\theta',t')$ denote the same line integral, it is obvious that:

$$
\frac{\partial}{\partial t}Rf(\theta,t) = \frac{\partial}{\partial t'}Rf(\theta',t'). \tag{7}
$$

Based on Eqs. (6) and (7), we have:

$$
\begin{aligned}
&\frac{\partial}{\partial\theta'}Rf(\theta',t') - \frac{\partial}{\partial\theta}Rf(\theta,t) = \tau\frac{\partial}{\partial t}Rf(\theta,t), \\
&\text{or, } \frac{\partial}{\partial t}Rf(\theta,t) = \frac{1}{\tau}\left(\frac{\partial}{\partial\theta'}Rf(\theta',t') - \frac{\partial}{\partial\theta}Rf(\theta,t)\right).
\end{aligned}
\tag{8}
$$

The given equation states that the partial derivative of the Radon transform with respect to the distance can be obtained by taking the difference of the partial derivatives with respect to the projection angles at two different rotation centers. This relationship is evidently independent of the imaging object because the equation is just based on the Radon transform $Rf(\theta,t)$ no matter what the object $f$ is. Hence, Eq. (8) represents a universal local correlation equation of PDE, abbreviated as LCE.

To further derivation, we simplified Eq.(8) into following form:

$$
\frac{\partial R}{\partial t} = \frac{1}{\tau}\left(\frac{\partial R}{\partial\theta'} - \frac{\partial R}{\partial\theta}\right), \tag{9}
$$

$$
\frac{\partial R}{\partial\theta} + \tau\frac{\partial R}{\partial t} = \frac{\partial R}{\partial\theta'}\ . \tag{10}
$$

To deduce higher order LCE, taking the $\frac{\partial}{\partial\theta'}$ of both side, Eq.(10) becomes:

$$
\frac{\partial^2 R}{\partial\theta\partial\theta'} + \tau\frac{\partial^2 R}{\partial t\partial\theta'} = \frac{\partial^2 R}{\partial\theta'^2}\ . \tag{11}
$$

Substituting Eq. (10) into Eq.(11), we have the 2nd order LCE:

$$
\frac{\partial^2 R}{\partial\theta^2} + 2\tau\frac{\partial^2 R}{\partial t\partial\theta} + \tau^2\frac{\partial^2 R}{\partial t^2} = \frac{\partial^2 R}{\partial\theta'^2}\ . \tag{12}
$$

Based on Eq.(12), we can deduce the 3rd order LCE of $\theta$:

$$
\frac{\partial^3 R}{\partial\theta^2\partial\theta'} + 2\tau\frac{\partial^3 R}{\partial t\partial\theta\partial\theta'} + \tau^2\frac{\partial^3 R}{\partial t^2\partial\theta'} = \frac{\partial^3 R}{\partial\theta'^3}\ . \tag{13}
$$

Substituting Eq.(10) into Eq.(13), the 3rd order LCE can be written as:

$$\frac{\partial^2}{\partial\theta^2}\left(\frac{\partial R}{\partial\theta}+\tau\frac{\partial R}{\partial t}\right)+2\tau\frac{\partial^2}{\partial t\partial\theta}\left(\frac{\partial R}{\partial\theta}+\tau\frac{\partial R}{\partial t}\right)+\tau^2\frac{\partial^2}{\partial t^2}\left(\frac{\partial R}{\partial\theta}+\tau\frac{\partial R}{\partial t}\right)=\frac{\partial^3 R}{\partial\theta'^3},$$

$$\frac{\partial^3 R}{\partial\theta^3}+\tau\frac{\partial^3 R}{\partial t\partial\theta^2}+2\tau\frac{\partial^3 R}{\partial t\partial\theta^2}+2\tau^2\frac{\partial^3 R}{\partial t^2\partial\theta}+\tau^2\frac{\partial^3 R}{\partial t^2\partial\theta}+\tau^3\frac{\partial^3 R}{\partial t^3}=\frac{\partial^3 R}{\partial\theta'^3},$$

$$\frac{\partial^3 R}{\partial\theta^3}+3\tau\frac{\partial^3 R}{\partial t\partial\theta^2}+3\tau^2\frac{\partial^3 R}{\partial t^2\partial\theta}+\tau^3\frac{\partial^3 R}{\partial t^3}=\frac{\partial^3 R}{\partial\theta'^3}. \tag{14}$$

In the same way, the 4th order LCE can be formulized as:

$$\frac{\partial^4 R}{\partial\theta^4}+4\tau\frac{\partial^4 R}{\partial t\partial\theta^3}+6\tau^2\frac{\partial^4 R}{\partial t^2\partial\theta^2}+4\tau^3\frac{\partial^4 R}{\partial t^3\partial\theta}+\tau^4\frac{\partial^4 R}{\partial t^4}=\frac{\partial^4 R}{\partial\theta'^4}. \tag{15}$$

For simplicity, we omit the further derivation for higher order LCEs. In summary, the coefficients of the above LCE conform to the Pascal's triangle relationship. Hence, any order LCE can be written with the binomial coefficients in the following form:

$$\sum_{k=0}^{n}\frac{n!}{k!(n-k)!}\frac{\partial^n R}{\tau^k\partial t^{n-k}\partial\theta^k}=\frac{1}{\tau^n}\frac{\partial^n R}{\partial\theta'^n} \tag{16}$$

The proposed LCE is built on the CT geometry of dual centers of rotation, which can be directly applied to divergent beam CT geometries, such as fan beam or cone beam CT systems, by setting one rotation center, i.e., $\theta'$, on the focal spot $\xi$. In the fan beam CT or any 2-D profile of cone beam CT, the angular derivate of Radon transform on the focal spot is available by calculating the derivate with respect to detector. In this case, the LCE shown in Eq.(16) demonstrates the general correlation property of the CT projections over neighbouring coordinates of $(\theta,t)$. In the nature of that the LCE is independent to imaging object, the LCE can be used to solve various problems existed in current CT systems by exploring the redundancy of projection data, such as sparse-view CT and denoise in low-dose projection data.

In the following subsections, we will give two equivalent LCEs under different fan beam trajectories.

### *C. Equivalent LCE based on circular scanning trajectory*

Circular scanning trajectory is a widely used geometry in clinical CT and most other CT devices. In this subsection, we consider an equiangular fan-beam geometry, as shown in Fig. 3. In this geometry, all the integral lines insect at the point $\xi$. We set the second rotation center $\mathbf{O}'$ at the point $\xi$. Correspondingly, $\theta'$ can be replaced by $\omega$, and $\tau = s$. It is worth noting that $s$ varies according to $\theta$ as $s = s_0\cos\theta$, while $s_0$ is the distance of source to iso-center. Hence, Eq.(16) becomes:

$$\sum_{k=0}^{n}\frac{n!}{k!(n-k)!}\frac{\partial^n R}{s^k\partial t^{n-k}\partial\theta^k}=\frac{1}{s^n}\frac{\partial^n R}{\partial\omega^n}. \tag{17}$$

We name Eq.(17) as cCLE (CLE on circular scanning trajectory). Specifically, the 1st and 2nd order cLCEs are as follows:

$$\frac{\partial R}{\partial t}=\frac{1}{s}\left(\frac{\partial R}{\partial\omega}-\frac{\partial R}{\partial\theta}\right) \tag{18}$$

$$\frac{\partial^2 R}{\partial t^2}+\frac{2}{s}\frac{\partial^2 R}{\partial\theta\partial t}+\frac{1}{s^2}\frac{\partial^2 R}{\partial\theta^2}=\frac{1}{s^2}\frac{\partial^2 R}{\partial\omega^2}, \tag{19}$$

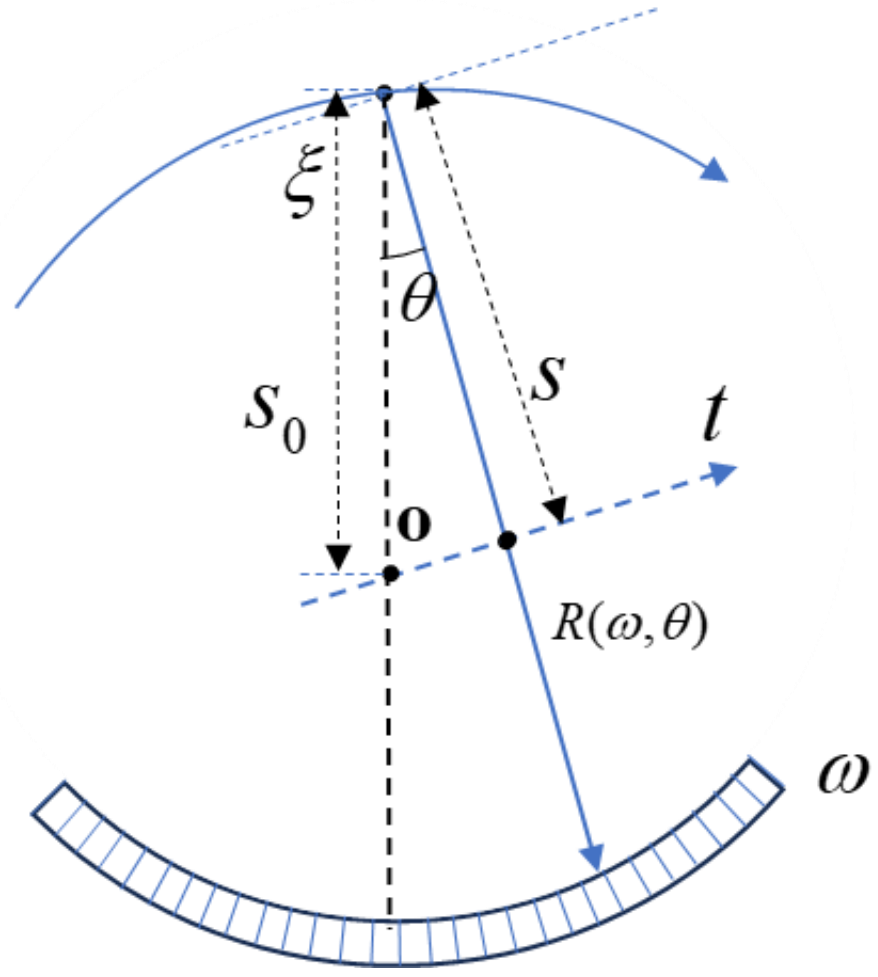

Fig. 3. The dual centers of rotation in circular trajectory

Especially, for the LCE along the line from $\xi$ to $\omega$, when we move the rotation center $\mathbf{O}$ to the coordinate of $\omega$, a specific equation can be generated as:

$$\frac{\partial R}{\partial t}=\frac{1}{s}\left(\frac{\partial R}{\partial\omega}-\frac{\partial R}{\partial v}\right) \tag{20}$$

$$\frac{\partial^2 R}{\partial t^2}+\frac{2}{s}\frac{\partial^2 R}{\partial v\partial t}+\frac{1}{s^2}\frac{\partial^2 R}{\partial v^2}=\frac{1}{s^2}\frac{\partial^2 R}{\partial\omega^2}, \tag{21}$$

where the derivates of R with regard to v denote the derivates from adjacent sources with same detector index. In this study, we will use a discrete form of the first order PDE in Eq.(20) to propose a cLCE based interpolation method.

### *D. Equivalent LCE based stationary linear array scanning trajectory*

With the emergence of the field emission cold cathode X-ray source, line distributed stationary scanning geometry has been gradually employed in CT imaging, such as digital breast tomosynthesis, etc (Yang *et al.*, 2008; Inscoe *et al.*, 2022). With CNT, researchers are able to design novel stationary geometry and save scanning time. In this subsection, we modify the LCE to cope with the line distributed source configuration with the linear array detector.

In the line distributed source configuration shown in Fig. 4, we introduce the auxiliary variables $o$, $d$ to cope with the linear array detector. Let $d$ denote the distance between pseudo rotation center and the line $\eta$. It is easy to find that:

$$\begin{aligned}\partial t &= \cos\theta\partial o\\ d &= t\cos\theta.\end{aligned} \tag{22}$$

The linear array detector geometry is also available to cone beam CT scenario by resolving any 2-D profile of axial view of the 3-D cone beam projection. For simplicity, we also set an auxiliary variable of equiangular detector $\theta'$, which intersects with the detector on $\eta$. And the distance between source and detector is $H$, as shown in Fig. 4. We set the dual centers of rotation at $\xi$ and the intersected point of the integral line with the axial $o$ , respectively. Based on the relationship, we can conclude any order equivalent LCEs, namely sLCEs, of stationary linear array source with linear array detectors as:

$$\sum_{k=0}^{n}\frac{n!}{k!(n-k)!}\frac{\cos^{2k}\theta}{(H-d)^k}\frac{\partial^n R}{\partial o^{n-k}\partial\theta^k}=\frac{\cos^{2n}\theta}{(H-d)^n}\frac{\partial^n R}{\partial\theta'^n}. \tag{23}$$

According to the relationship between $\eta$ and $H$ :

$$\eta = H\tan\theta', \tag{24}$$

the derivates of $\frac{\partial R}{\partial\theta'},\frac{\partial^2 R}{\partial\theta'^2}$ can be calculated as:

$$\begin{aligned}\frac{\partial R}{\partial\theta'}&=\frac{H}{\cos^2\theta}\frac{\partial}{\partial\eta}\\ \frac{\partial^2 R}{\partial\theta'^2}&=\frac{2H\sin\theta}{\cos^3\theta}\frac{\partial R}{\partial\eta}+\frac{H^2}{\cos^4\theta}\frac{\partial^2 R}{\partial\eta^2}.\end{aligned} \tag{25}$$

Substituting Eq.(25) into Eq.(23), the 1st sLCE becomes:

$$\frac{\partial R}{\partial o}=\frac{\cos^2\theta}{d-H}\left(\frac{\partial R}{\partial\theta}-\frac{H}{\cos^2\theta}\frac{\partial R}{\partial\eta}\right), \tag{26}$$

and the 2nd sLCE becomes:

$$\begin{aligned}&\frac{\partial^2 R}{\partial o^2}-\frac{2\cos^2\theta}{(d-H)}\frac{\partial^2 R}{\partial\theta\partial o}+\frac{\cos^4\theta}{(d-H)^2}\frac{\partial^2 R}{\partial\theta^2}=\\ &\frac{2H\sin\theta\cos\theta}{(d-H)^2}\frac{\partial R}{\partial\eta}+\frac{H^2}{(d-H)^2}\frac{\partial^2 R}{\partial\eta^2}\end{aligned} \tag{27}$$

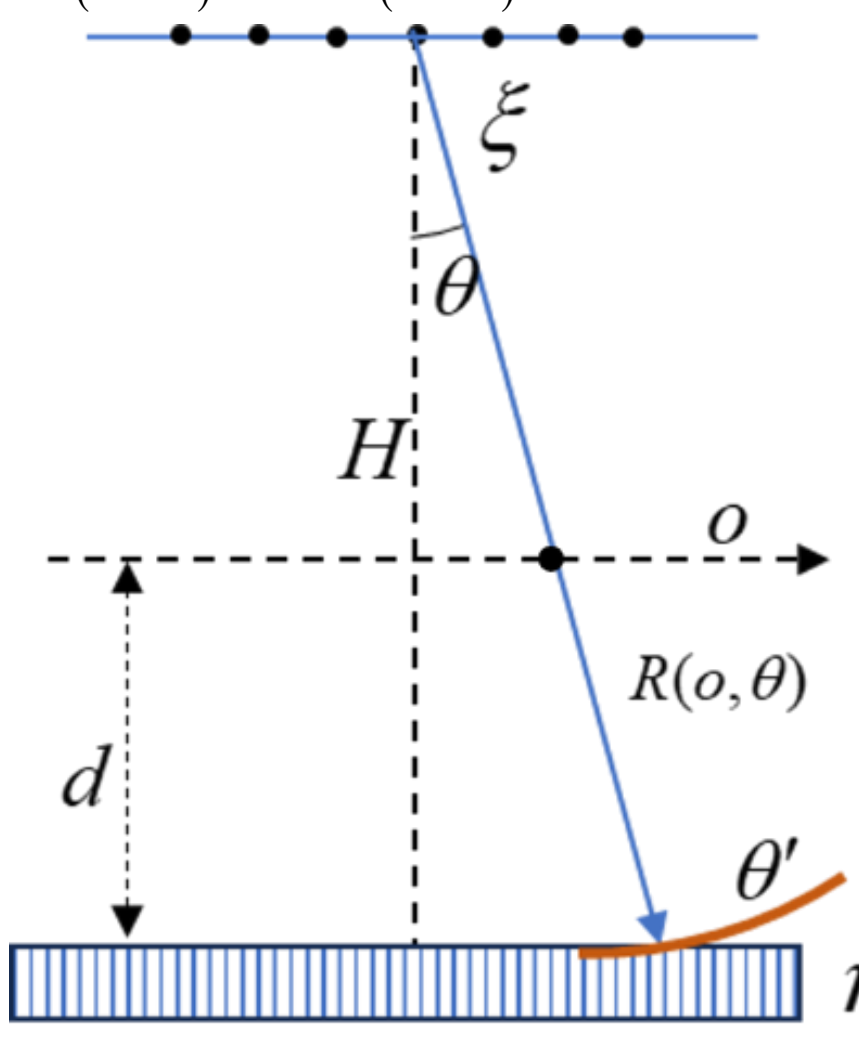

Fig. 4. The dual centers of rotation in stationary linear array scanning trajectory

## *E. Validation of the equivalent LCEs*

To validate the correctness of the proposed formula, we conducted a series of simulated experiments. We performed discrete calculations on both sides of the equation to determine the correctness of proposed LCEs. In simulation, XCAT (Segars *et al.*, 2010) was chosen as the simulation phantom. A series of XCAT slices were chosen to validate the proposed LCEs. Here, we list slice 200 of the XCAT (bottom left of Fig. 5) as the example to show the validation result. In circular scanning trajectory, the geometry settings are listed in Table 1. For validation of stationary linear array trajectory, the image volume size is $1.2\text{mm}\times1.2\text{mm}$ . Meanwhile, the distance between source and detector is $H=800mm$ and $d=50mm$ . The detector size is $1\text{mm}\times1\text{mm}$ .

Figs.5 and 6 depict the first order and second order cLCEs of the circular scanning trajectory, respectively. From the results, we can find that the discrete calculation on the left side gives almost the same value as calculated by the right side, except a little of error that are caused by discretization in numerical calculation. Therefore, we can conclude that the equation is correct for the given discrete values. Similar conclusion can be found in Figs.7 and 8.

Table 1. the circular scan trajectory settings

| | |
|---|---|
| Source to rotation center | 595mm |
| Field of view (FOV) | 500 |
| Detector number | 673 |
| Max angle | 26.6° |
| Min angle | -26.6° |
| Full of Views | 1164 |

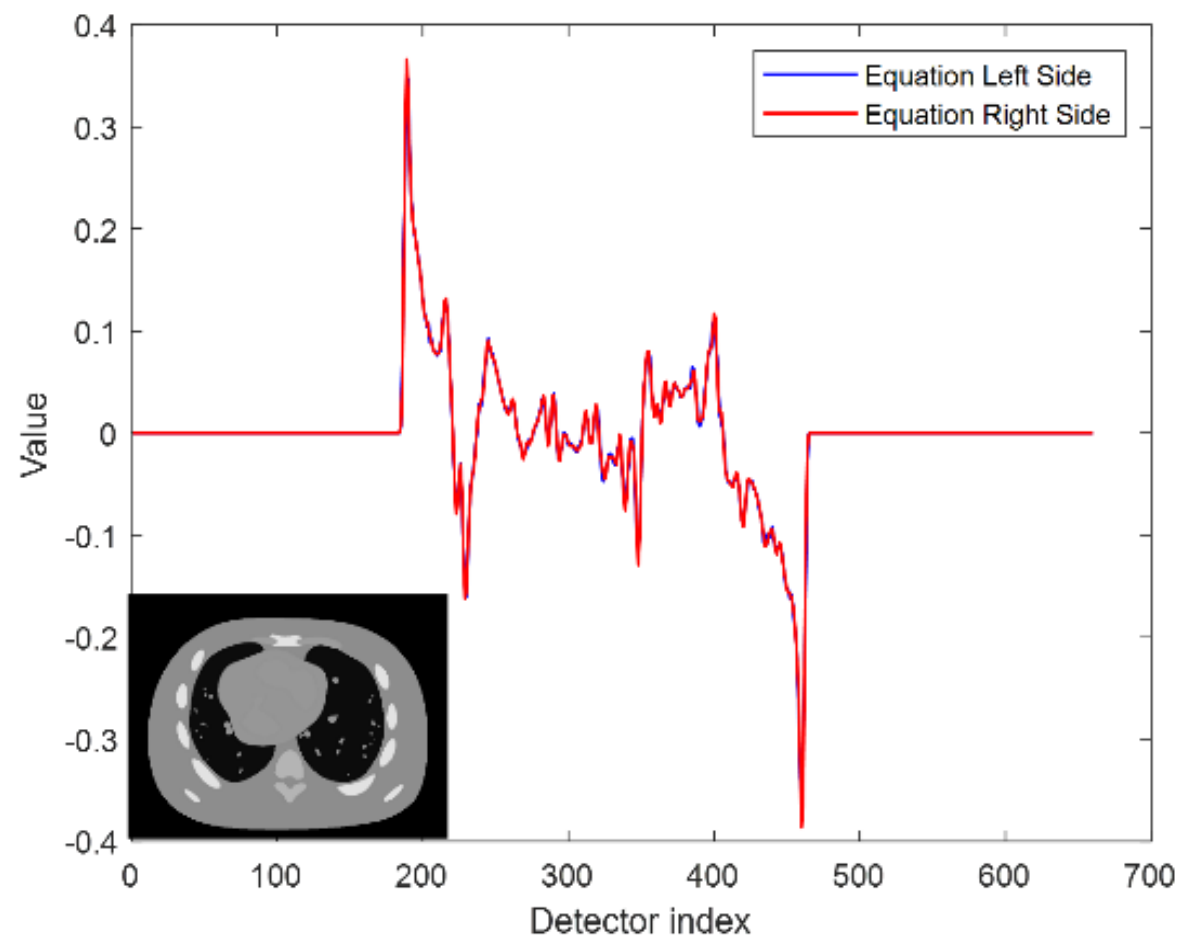

Fig. 5. The validation result of the first order cLCE based on the circular scanning trajectory (Projection angle = 0°).

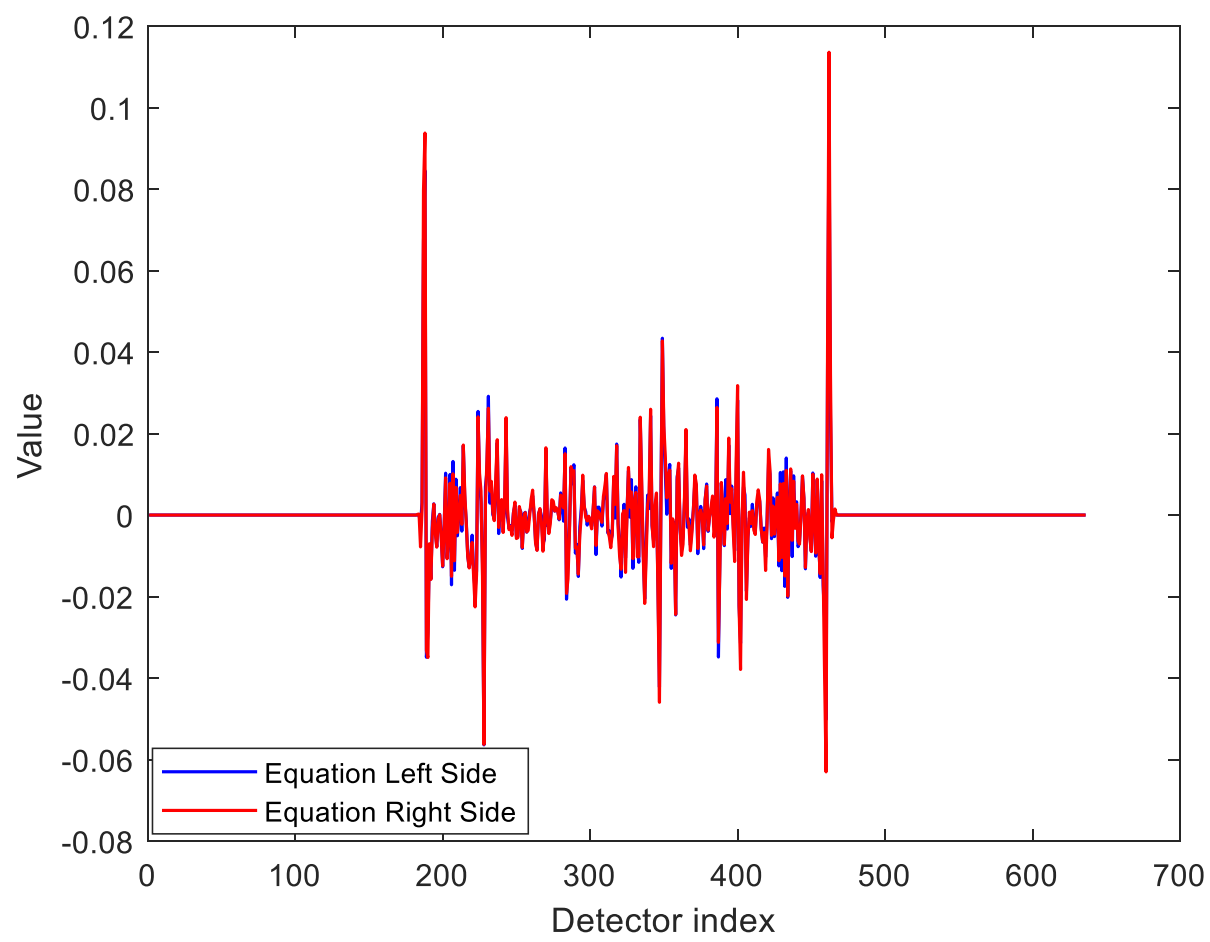

Fig. 6. The validation result of the second order cLCE based on the circular scanning trajectory (Projection angle = 0°).

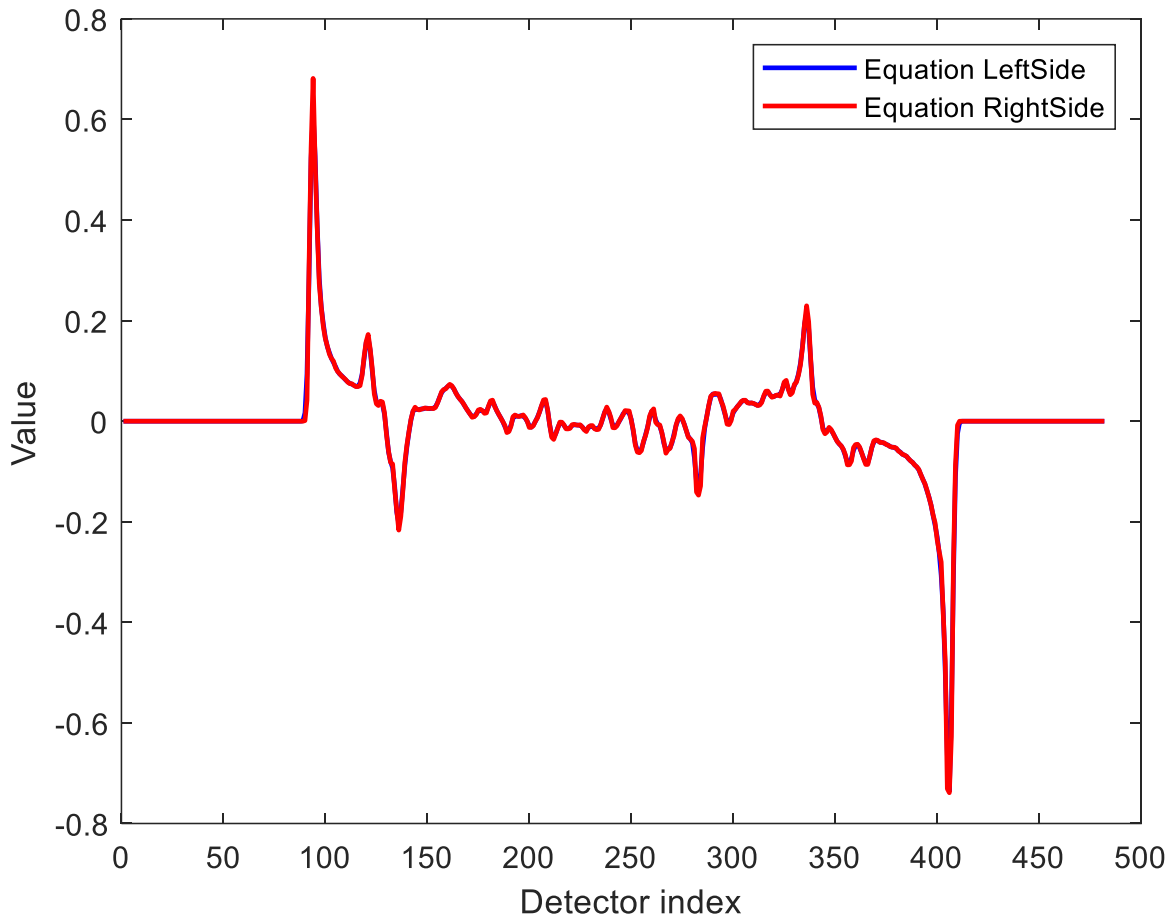

Fig. 7. The validation results of the first order sLCE based on the stationary linear array scanning trajectory.

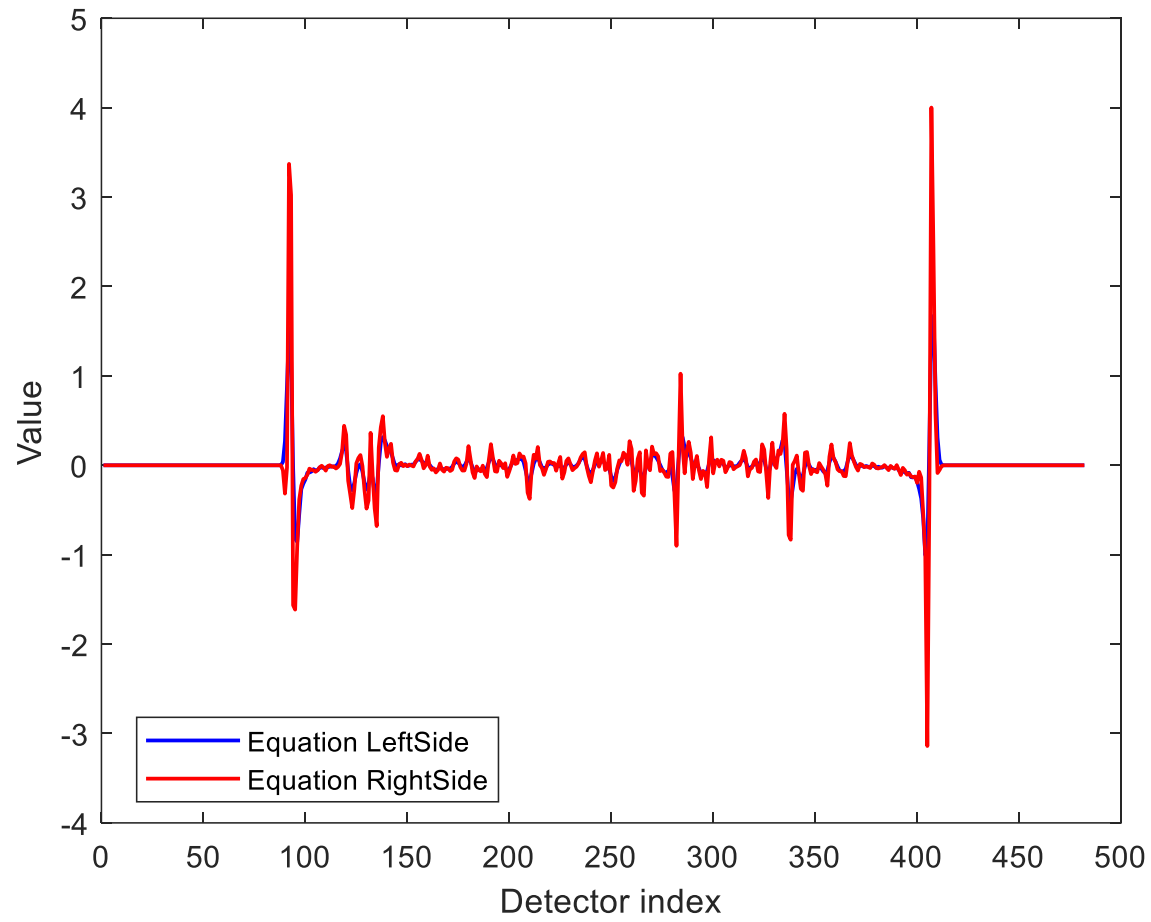

Fig. 8. The validation results of the second order sLCE based on the stationary linear array scanning trajectory.

## 3. Sparse-view reconstructions using the cLCE

In former section, we have showed the correctness of the cLCEs and sLCEs. In this section, we will employ the cLCE to solve sparse-view problem of circular scanning trajectory, since the circular geometry is widely used in current CT equipment.

In this study, we would like to take a simplest scheme to achieve the sparse-view restoration. By using the simplest scheme that provides a least bound of interpolation ability, we will realize the potential of the cLCE based projection interpolation. To this end, we use a discrete form of the first order cLCE shown in Eq.(20) to interpolate the missing data.

To illustrate the interpolation scheme, firstly, we will detail the data configuration in circular scanning trajectory shown in Fig. 9. To accord with the proposed cLCE, we set to recover fan beam projections of the virtual sources with circular angle interval of $\Delta\omega$, where $\Delta\omega$ is the fan beam circular angle interval. Such the projection of all the virtual sources is called as complete view in this study. Let $M$ denote the complete view number, $M = \pi / \Delta\omega$. The complete view is dense and its projections are overcomplete. In this study, for the studied example of SIEMENS Somatom Sensation 64-slice CT scanner, the angular interval of scanning views is about $2\Delta\omega$, which can guarantee precise reconstruction with sufficient number of projection views. In this case, we realize any projection interpolation via recovering the complete view. In specific, we resolve a sparse-view CT problem by interpolating among the sparse projection to achieve a clinical complete projection via recovering the complete view. Let N represent the total detector number. And the detector index can be calculated as:

$$n' = n - \left\lceil \frac{N}{2} \right\rceil, n = 0,1,...,N-1. \tag{28}$$

Let $R_{m,n}$ denote the line integral between source m and detector n, as Fig. 9 shows. Based on the first order cLCE in Eq.(20), we chose six discrete projections to calculate the derivatives. A group of the six projections is exampled in Fig. 9, as marked by coloured rays. The projections are indexed by considering the geometric relationship among each ray of each source: $R_{m-1,n+2}, R_{m+1,n-2}, R_{m,n-1}, R_{m,n+1}, R_{m-1,n+1}, R_{m+1,n-1}$, as shown in Fig. 9 and Table 2. Hence, we can write the discrete format of the Eq.(20) as:

$$\begin{aligned} & R_{m+1,n-2} - R_{m-1,n+2} \\ & -\cos(n'\Delta\omega)(R_{m+1,n-1} + R_{m,n-1} - R_{m-1,n+1} - R_{m,n+1}) = 0. \end{aligned} \tag{29}$$

We can also write Eq.(29) in matrix form, as Eq.(30) lists. Here we construct a cLCE projection matrix $\mathbf{P} \subseteq R^{MN \times MN}$ to calculate the cLCE for whole projections $\mathbf{R} \subseteq R^{MN \times 1}$, with each row ( $m \times M + n,\ m \in [0, M-1],\ n \in [0, N-1]$ ) of $\mathbf{P}$ corresponding to Eq. (29), and $\mathbf{R}$ stacking all the projections of $M$ sources and $N$ detector elements for each source.

Obviously, we have:

$$\mathbf{PR} = \mathbf{0}. \tag{30}$$

To solve the PDE shown in Eq.(30), we should have the boundary condition that are the measured sparse-view projections. As an implementation, we consider the known sparse-view projection as constraint (Li *et al.*, 2019). Hence, the missing data in the sparse view projection can be solved by optimizing Eq.(31):

$$\arg\min_{\mathbf{R}} \{ \|\mathbf{PR}\|_2^2 + \alpha_1 \|\mathbf{D}^s \odot \mathbf{R} - \mathbf{R}^s\|_2^2 \}, \tag{31}$$

where $\mathbf{D}^s \subseteq R^{MN \times 1}$ describes the known projection locations, which is a sparse matrix with the elements corresponding to the known projection equal to 1, and the unmeasured elements equal to 0. $\odot$ represents the Hadamard production. $\mathbf{R}^s \subseteq R^{MN \times 1}$ denotes the known sparse view projection, with each of the unmeasured projections equals to $0$.

The solution of Eq. (31) is an ill-problem. To handle the problem, many optimization methods are available, e.g., ridge regression (Hoerl and Kennard, 1970) or Lasso (Santosa and Symes, 1986). However, in this study, we prefer to directly resolve the Eq. (31) without any extra process for the purpose to verify the projection interpolation ability by the proposed cLCE, by providing a least bound of the interpolation ability. In this study, Eq. (31) was resolved by gradient descent (Landweber, 1951).

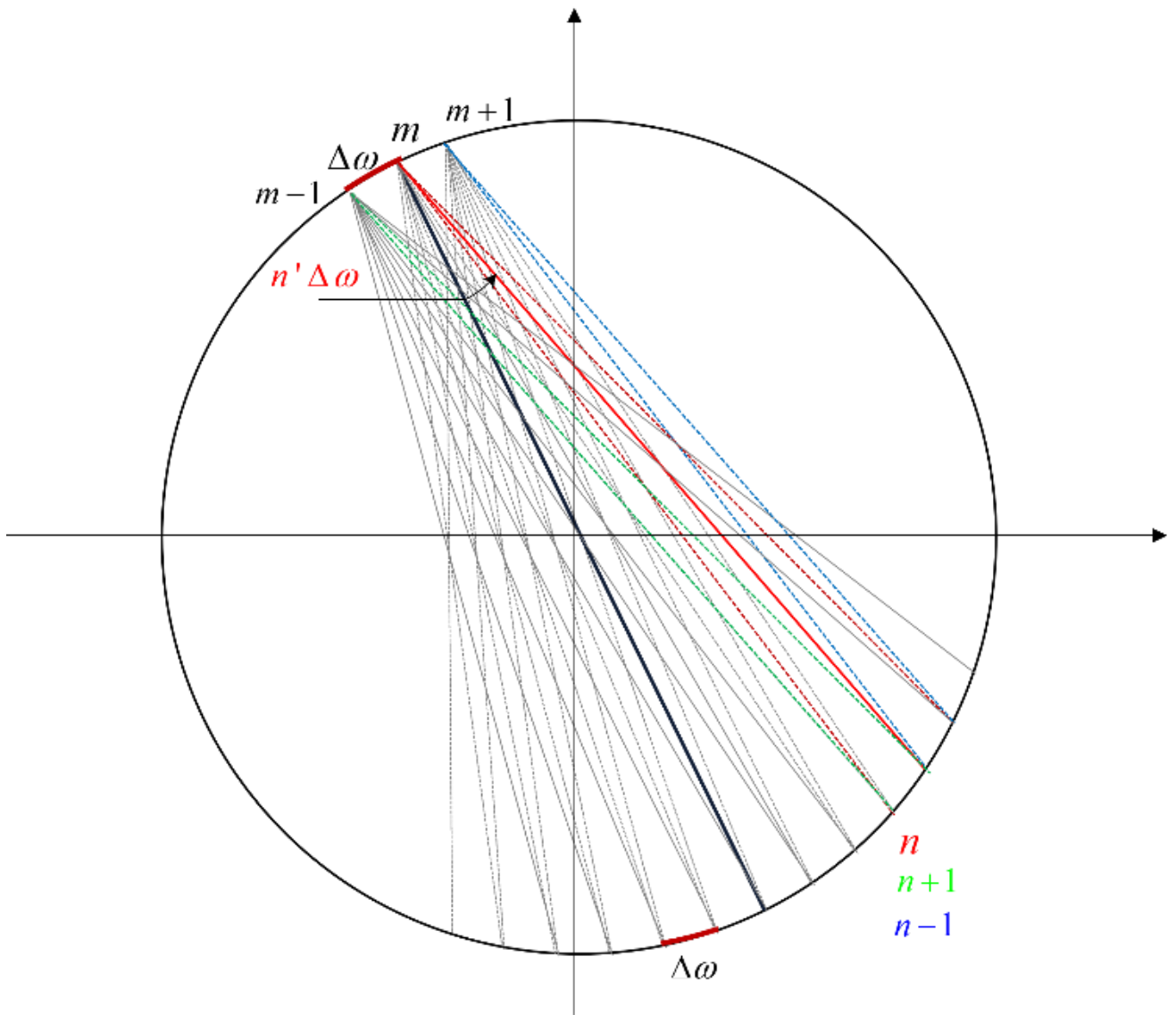

Fig. 9. Example to illustrate the interpolation procedure in circular scanning trajectory. Dotted coloured lines represent chosen rays for interpolation regard to red line. Red arc denotes the $\Delta\omega$ interval in complete view.

Table 2 the search table configuration

| Source Index | -3 | -2 | -1 | 0 | 1 | 2 | 3 |
|---|---|---|---|---|---|---|---|
| m-2 | n-1 | n | n+1 | n+2 | n+3 | n+4 | n+5 |
| m-1 | n-2 | n-1 | n | n+1 | n+2 | n+3 | n+4 |
| m | n-3 | n-2 | n-1 | n | n+1 | n+2 | n+3 |
| m+1 | n-4 | n-3 | n-2 | n-1 | n | n+1 | n+2 |
| m+2 | n-5 | n-4 | n-3 | n-2 | n-1 | n | n+1 |

The cLCE is of a PDE that is sensitive to noise or error. Moreover, discrete implementation also inevitably bring error in calculation. To suppress the noise and error during the interpolation, one possible way is to combine the interpolation into a regularized iterative reconstruction. Under this circumstance, the noise and error in the discrete cCLE based interpolation can be restrained by image regularization in iterative reconstruction. To this regard, we propose a unified reconstruction scheme to obtain the reconstructed image and projection interpolation simultaneously, as formularized by the following multi-objective optimization:

$$\arg\min_{\mathbf{x},\mathbf{R}}\{\|\mathbf{PR}\|_2^2+\alpha_1\|\mathbf{Ax}-\mathbf{R}\|_2^2+\alpha_2\|\mathbf{D}^s\odot\mathbf{R}-\mathbf{R}^s\|_2^2+\alpha_3\|\mathbf{A}_s\mathbf{x}-\mathbf{R}_s\|_2^2+\beta TV(\mathbf{x})\}. \tag{32}$$

Except the notations defined above, $\mathbf{A},\mathbf{A}_s$ represent system matrix and sparse sampled system matrix, respectively. $TV(.)$ is the total variation regularization term. $\alpha_i, i=1,2,3$ represent the regularization parameters. $\beta$ is TV regularization parameter. With the help of Eq.(32), the performance of interpolating higher sparse view problem can be guaranteed. In this study, this optimization function was divided into two subproblems(Xu and Yin, 2013):

$$\begin{aligned} &sub_1:\min_{\mathbf{R}}\|\mathbf{PR}\|_2^2+\alpha_2\|\mathbf{D}^s\odot\mathbf{R}-\mathbf{R}^s\|_2^2 \\ &sub_2:\min_{\mathbf{x}}\alpha_1\|\mathbf{Ax}-\mathbf{R}\|_2^2+\alpha_3\|\mathbf{A}_s\mathbf{x}-\mathbf{R}_s\|_2^2+\beta TV(\mathbf{x}). \end{aligned} \tag{33}$$

The final solution can be obtained through alternating iterations. To speed up the iterative process, we linearly interpolated the sparse projection to generate a full-angle projection as $\mathbf{R}$ initialization.

## 4. Reconstructions and analysis

Before detailing the experiment settings, we first clarify the relationship among complete view, clinical complete view, sparse view, and interpolated view. In this study, as mentioned above, complete view refers to the fan beam projections of virtual sources with circular angle interval of $\Delta\omega$ that is also the fan beam angular interval, as shown in Fig. 9. In the exampled clinical CT of this study, the scanning angular interval is normally $2\Delta\omega$ , which can guarantee perfect reconstruction with sufficient view number. We denote the projections as clinical complete view. Moreover, we follow the traditional definition of sparse view, in which sparsity level is relative to the clinical complete view. Remind again that our proposed method is to interpolate the sparse-view projection via recovering the complete view. More details about the experiment setting for the interpolation can be found in section 4C.

### *A. Phantoms used in validation*

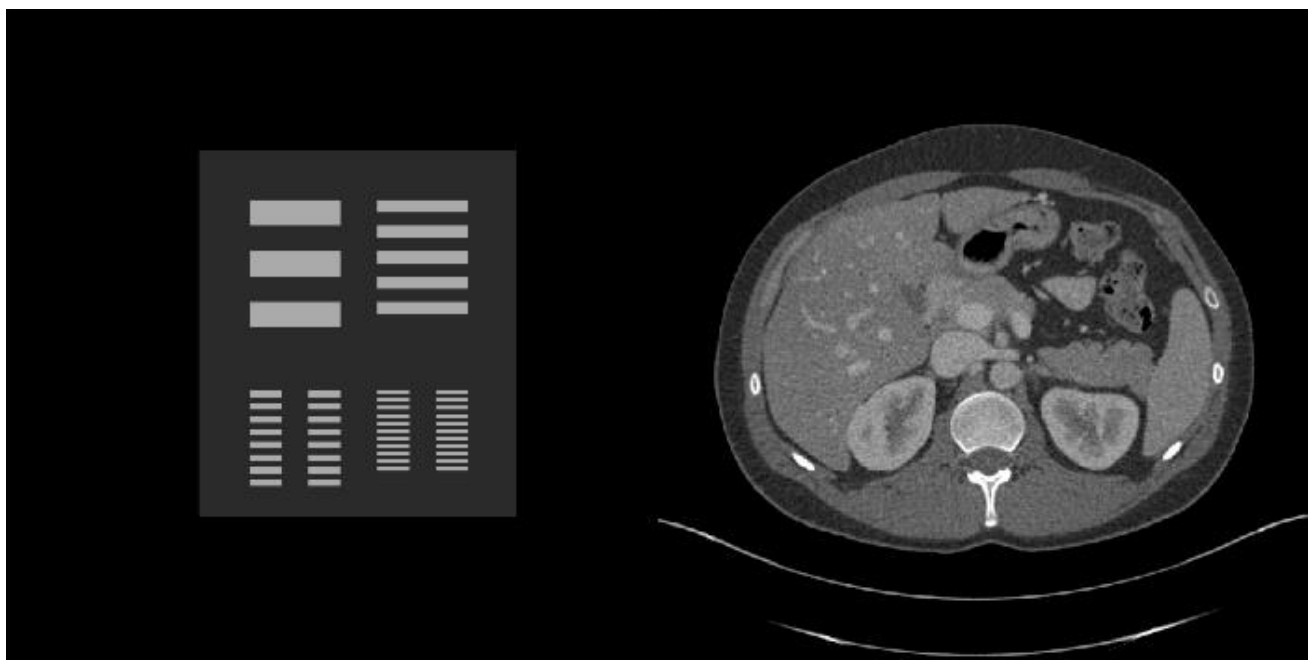

Fig. 10. The reconstruction phantom used in interpolation strategy.

To evaluate the proposed interpolation scheme and the unified reconstruction method, we designed a simulation phantom with different resolution line pairs, which is shown in Fig. 10. The simulated line widths are 2 pixels, 3 pixels, 10 pixels and 20 pixels, respectively. While in real data validation, we chose a real abdomen data from Mayo Low dose challenge (www.aapm.org/GrandChallenge/LowDoseCT/), also shown in Fig. 10. Both image size is $512\times512$ . The geometry settings are accordance with Table 1. Two different sparsity levels were set to validate the efficiency of the proposed interpolation methods based on cLCE. For specific, we used 1/4 sparsity to validate the performance of Eq.(31), which means directly using LCE to solve sparse view problem. Meanwhile, we used 1/8 sparsity to validate the unified optimization scheme of Eq. (32). The details can be found in Table 3.

Table 3. the projection number settings in experiments

| Sparsity level | Number of Complete view | Number of Clinical complete view | Number of Sparse view |
|---|---|---|---|
| 1/4 | 2328 | 1164 | 291 |
| 1/8 | 2328 | 1164 | 146 |

## B. Comparison methods and criteria

In 1/4 sparsity validation, we compared the interpolated sinogram profile with clinical complete sinogram to visualize the efficiency of the cLCE interpolation. Meanwhile, we used filtered back projection (FBP) to reconstruct the interpolated sinogram to further investigate the image quality of the interpolated sinogram. To evaluate the proposed unified framework quantitatively and qualifiedly, we compared the proposed method with the traditional iterative reconstruction and filtered back projection (FBP) in 1/8 sparsity experiment. In this study, the selected iterative reconstruction scheme consists of an algebraic reconstruction technique (ART) with TV regularization, named ART+TV. The comparison is also an ablation study on the unified reconstruction scheme since we can distinguish the interpolation effect by observing the improvement of image quality led by the cLCE interpolation based on the traditional ART+TV reconstruction.

In comparing the results based on the simulation phantom, we calculated MSE (mean square error) between ground truth (GT) images and the reconstructed results:

$$MSE(\mathbf{x},\mathbf{y}) = \frac{1}{N}\sum_{i=1}^{N}(\mathbf{x}^i - \mathbf{y}^i)^2, \tag{34}$$

where $\mathbf{x},\mathbf{y}$ represent the ground truth image and reconstructed image, respectively, $N$ is the total pixel number of the image. Moreover, we also drew a line profile to visualize the image quality of each reconstruction method.

In the experiment on Mayo abdomen data, we chose SSIM and CNR as evaluation metrics. The calculation of SSIM and CNR can be formularized as below:

$$\mathrm{SSIM}(\mathbf{x},\mathbf{y}) = \frac{(2\mu_x\mu_y + c_1)(2\delta_{xy} + c_2)}{(\mu_x{}^2 + \mu_y{}^2 + c_1)(\delta_x{}^2 + \delta_y{}^2 + c_2)}, \tag{35}$$

where $\mu_x, \mu_y$ are the mean of the reconstructed image from the clinical view projection and the evaluated image, and $\delta_x{}^2$, $\delta_y{}^2$, $\delta_{xy}$ are the variances and covariance of the two images, respectively. Also, $c_1, c_2$ are two small positive constants to stabilize the division with a weak denominator.

The formula of CNR is as below:

$$CNR = \frac{2\,|\,ROI_1 - ROI_2\,|}{\delta_{ROI_1} + \delta_{ROI_2}}\ , \tag{36}$$

where $ROI_1$ and $ROI_2$ are the mean intensity of the selected areas that represent signal and background, respectively. $\delta_{ROI_1}$ and $\delta_{ROI_2}$ are the corresponding standard deviation value.

## C. Results and analysis

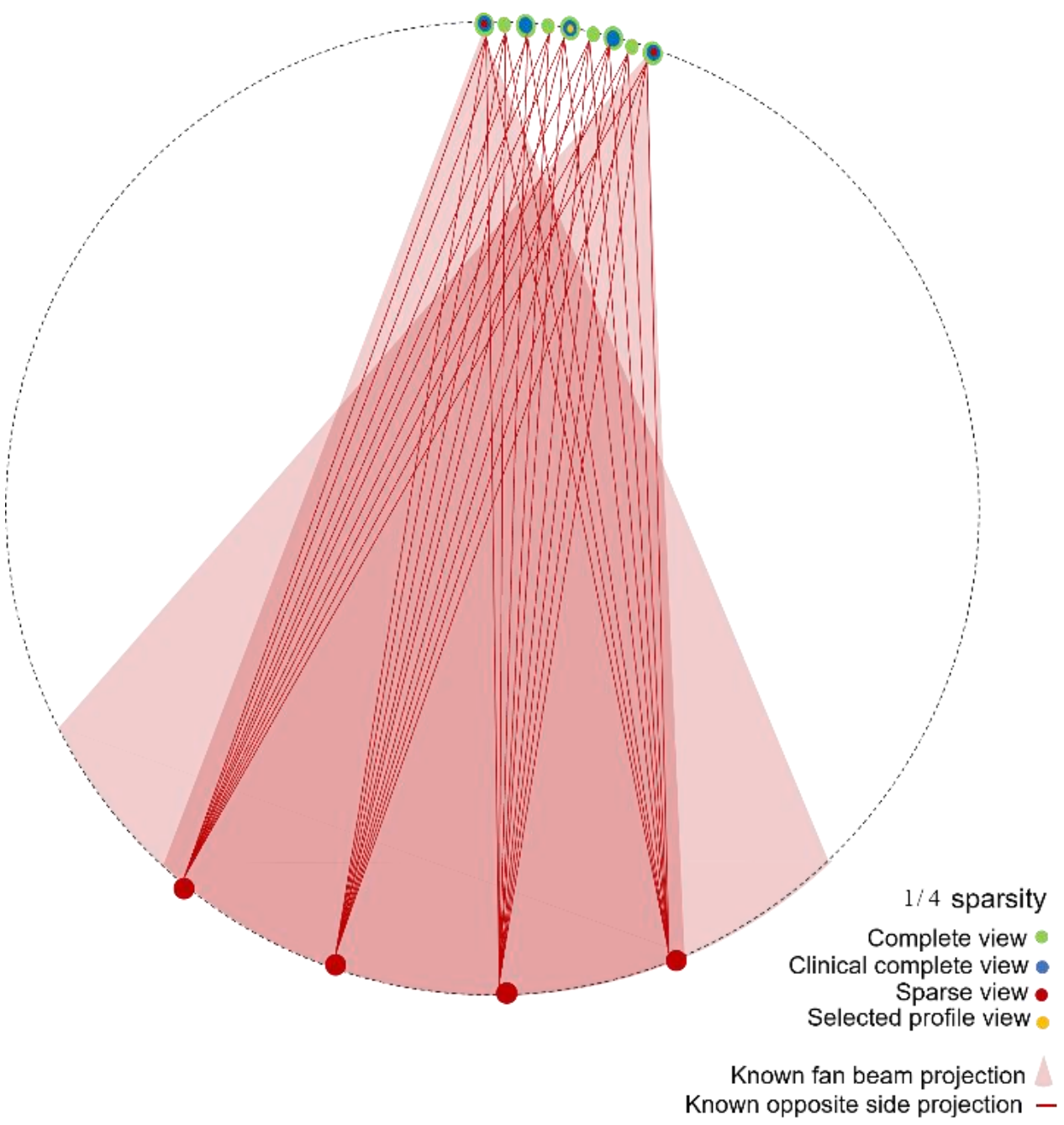

Fig. 11. The illustration of projection profile selection.

In this section, we took serval preliminary experiments to illustrate our proposed cLCE interpolation method. Before representing the results, we will first illustrate the projections for interpolation and comparison, as Fig. 11 shows. We took 1/4 sparsity as an example. Red dots denote the known sparse view sources. As mentioned before, the clinical complete view is not the complete view in this study. Hence, during the interpolation, we will recover the complete view in the scanning trajectory, as the beams sourced with green dots. The boundary value for the cLCE computation is the two sparse view projection that is denoted as red fan in Fig. 11. Besides, sparse samplings on the opposite side also provide a certain number of boundary conditions for interpolation, as indicated by red rays. Hence, the total boundary conditions are two adjacent sparse view projection and opposite sampling rays. After the complete view is recovered, the clinical complete view, as the beams indicated by blue dots, is also obtained. To visualize the efficiency of the cLCE, in profile visualization, we chose the center index of interpolated fan beam projection, as yellow dot denotes.

Figs.12 and 13 are the 1/4 sparsity interpolated results of the line pairs phantom. From left to right, the results of sparse view FBP, interpolated sinogram FBP, and the ground truth are presented, respectively. From the results, it is apparent that missing data in the sparse view projection can be well recovered by directly using the discrete first order cLCE. Via FBP process, it is shown that after LCE interpolation, the streak artefacts can be well eliminated without image detail degradation, as red arrows indicate. Fig. 13 shows two different interpolated projection profiles of different projection views. From the results, good agreement is found between the two profiles, which shows the cLCE equations can recover the complete view from the sparse view projection. Further, the picture depicts that the recovered projection is a little different from the GT value. We ascribe this phenomenon to twofold: 1) in this study, we only use the first order cLCE of angular derivates on circular trajectory, more redundancy regarding the adjacent rays has not been employed. 2) the discrete solution of the cLCE in this study inevitably brings errors, and derivate operator in the cLCE also amplifies the errors.

Figs.14 and 15 show the interpolated results of Mayo data. As evident, the cLCE can also capture the tendency of the small changes in the projection. The interpolated projections show the same tendency with the ground truth. Both cases show the efficiency of the cLCE. In Fig. 14, we chose an area indicated by a white dashed box to detail the difference among the different methods. The zoomed in maps are showed in second row in Fig. 14. It is evident that, compared with FBP, the artefacts in cLCE resolved result can be removed without detail loss, as red arrow indicated.

In validation of the unified reconstruction framework, we also employed resolution phantom and Mayo data as the imaging object and the sparsity level is 1/8. The results are shown in Figs.16-20. In each studied case, we optimized the parameters in resolving Eq. (32) to reach the best result. The final optimized parameters of each case can be found in Table 4.

Table 4. The parameters settings in each case

| | relax parameter | $\alpha_1$ | $\alpha_2$ | $\alpha_3$ | $\beta$ | Iteration number |
|---|---|---|---|---|---|---|
| Line pairs | 0.25 | 1 | 0.97 | 0.97 | 0.035 | 600 |
| Mayo | 0.2 | 1 | 0.97 | 0.97 | 0.019 | 700 |

Fig. 16 shows the interpolated results of the resolution phantom. From the results, it is observed that the recovered projection exhibits the same tendency as the GT. The result shows the correctness of the proposed interpolated strategy. Moreover, with the optimization, we can also obtain the reconstructed images shown in Fig. 17. From left to right, the reconstructed images of FBP, ART+TV, and the proposed method, as well as GT are illustrated, respectively. The MSE value was calculated and listed in the bottom of each reconstruction. From the results, it is shown that FBP exhibits severe streak artefacts which degrades the image quality. In ART+TV, the streak artefacts cannot be fully suppressed and a lot of image details are oversmoothed. For the proposed method, it is observed that while image details are well recovered, the streak artefacts are removed significantly, as red arrows indicate. This ablation experiment results indicate that even using the simplest interpolation scheme, i.e., the discrete first-order cLCE interpolation, the proposed method still achieves satisfied interpolation effect based on the ART+TV reconstruction for 1/8 sparse-view CT scenario. The MSE values also show the potential of the proposed method. To visualize the image quality by different reconstruction methods, we plotted line profile along yellow dashed line indicated in Fig. 17. Fig. 18 shows the line profiles of ART+TV, proposed method and GT, respectively. For better visualization, we enlarged the area denoted by yellow dashed box. It is apparent that proposed method has the best capacity to preserve details. The results also prove that the proposed method is able to simultaneously suppress streak artefacts and preserve details.

Figs.19 and 20 show the results of the abdomen data. The parameters used in this case can be found in Table 4. Fig. 19 shows the interpolated results of the abdomen data. As exemplified, the proposed method precisely predicts the absent projection from the sparse projection. Fig. 20 demonstrates the reconstruction results. In the first row, from left to right, the results of FBP. ART+TV, and the proposed method, as well as GT are showed, separately. In the second row, we calculated the residual images of each reconstruction. For better visualization, we chose a zoom in area indicated by red dashed box. The third row depicts the results of the enlarged area. As illustrated, the proposed method well recovers smaller details than ART+TV, as indicated by red arrows. At the same time, it also suppresses the severe streak artefacts compared with FBP. Besides, to quantitively evaluate the proposed method, the SSIM and CNR were calculated, which is listed in Table 5. The CNR areas are pointed in white and green boxes in Fig. 20. From the results, it indicates that the proposed method works well in preserving details and suppressing the artefacts at the same time, which shows the efficiency of the proposed method.

Table 5. SSIM and CNR values

| | FBP | ART+TV | Proposed | GT |
|---|---|---|---|---|
| SSIM | 0.70 | 0.9254 | 0.9318 | / |
| CNR | / | 73.03 | 77.55 | 78.98 |

To sum up, in this section, we validated the discrete cLCE interpolation method and the unified reconstruction scheme based on the cLCE. Both methods well recover the missing data from sparse view projection. The results based on the cLCE interpolation show that sparse view projection indeed consists of enough information of complete projection, which can be recovered by the proposed cLCE. In higher sparsity level, the unified reconstruction is more robust to cope with error and noise in handling the sparse view projection. To demonstrate the ability of the cLCE in the unified reconstruction framework, we involved an ablation experiment, as shown in Figs.17, 18 and 20. From the results, it is shown that the cLCE interpolation plays a vital role in suppressing artefacts and preserving details.

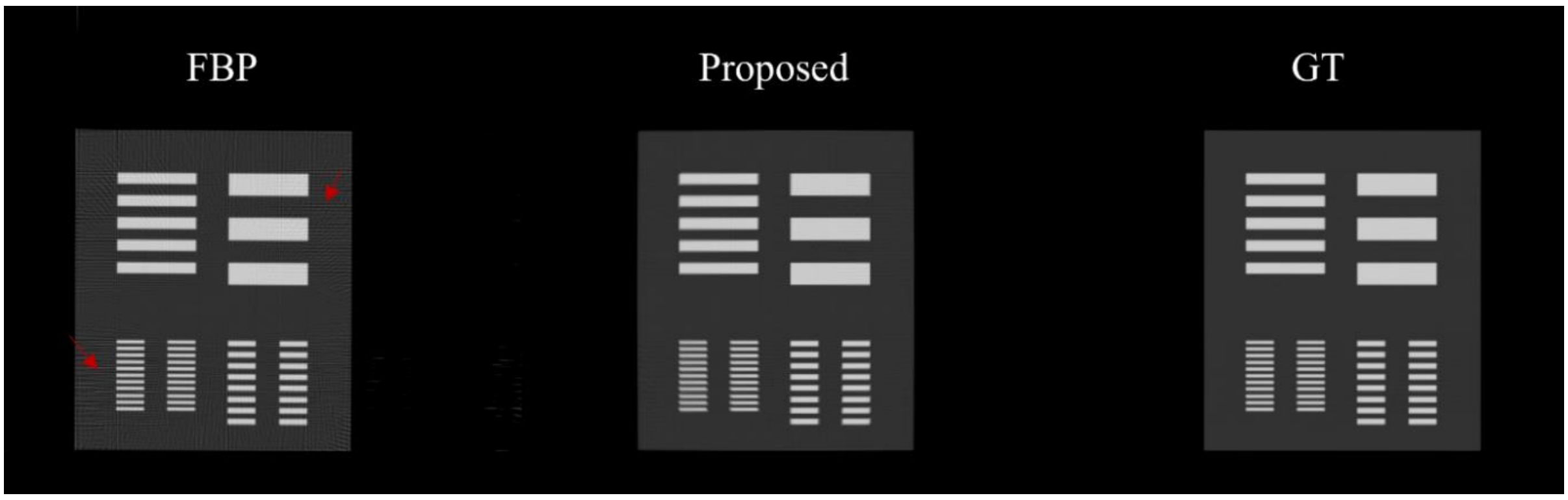

Fig. 12. 1/4 sparsity results of the resolution phantom. From left to right, it showed the sparse view FBP result, interpolated FBP result, ground truth, respectively.

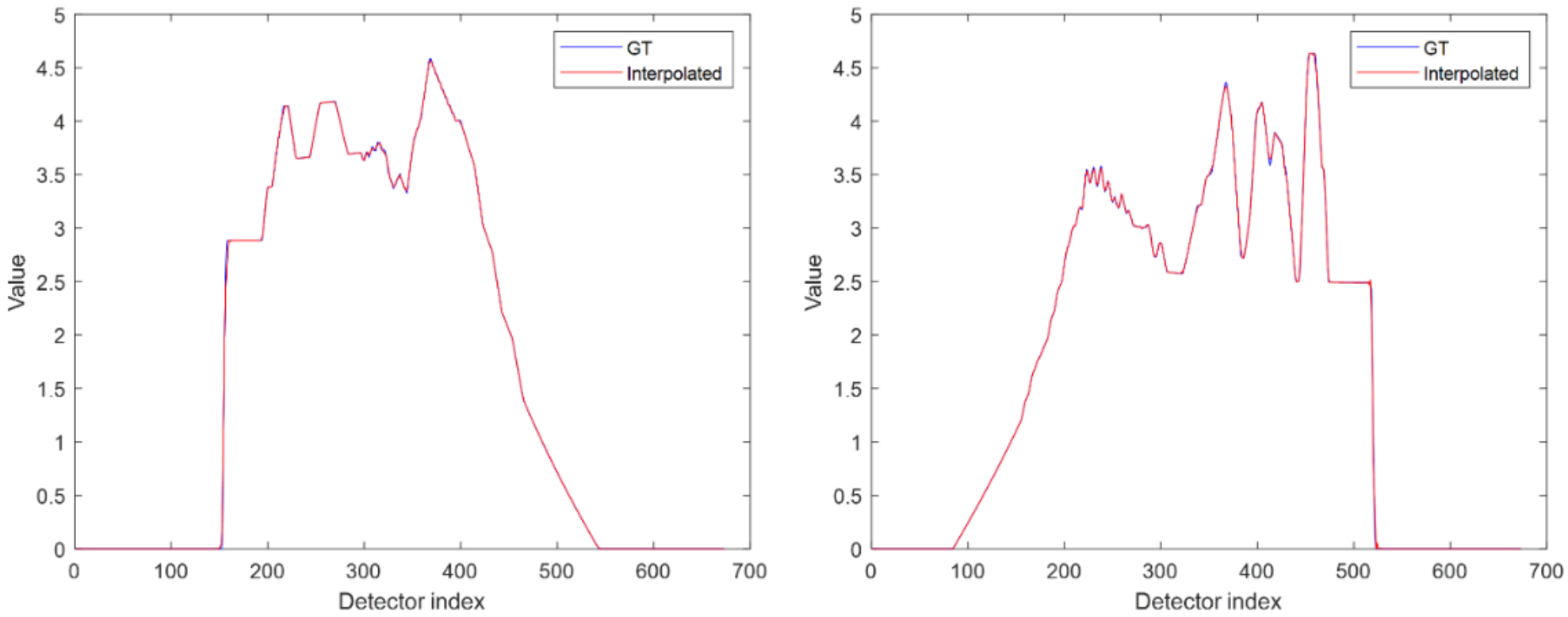

Fig. 13. 1/4 sparsity interpolation results of the resolution phantom. From left to right, it showed results of 5th and 805th projections.

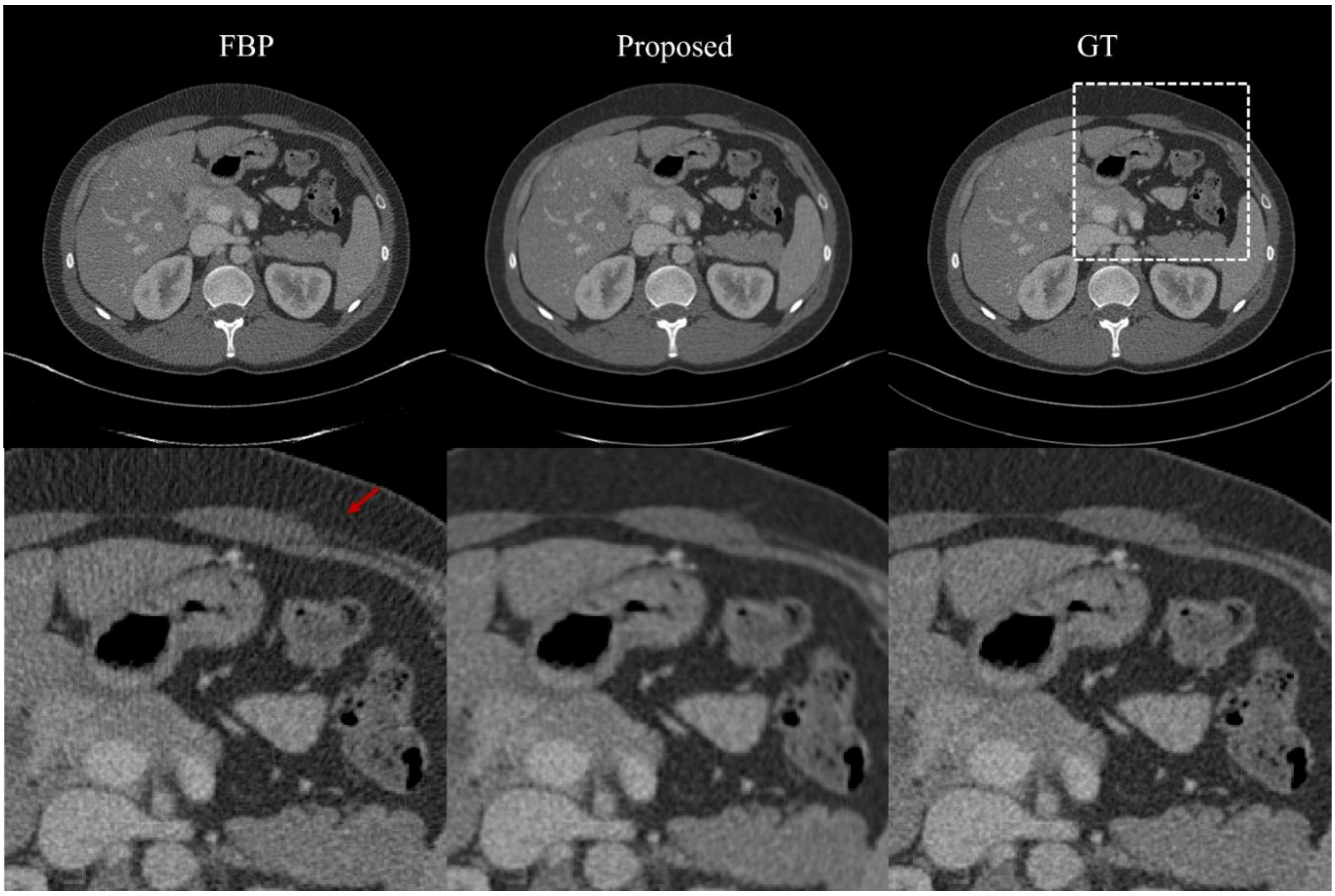

Fig. 14. 1/4 sparsity results of the Mayo data. In first row, from left to right, it showed the sparse view FBP result, interpolated FBP result, ground truth, respectively. Second row is the zoom in area in white dashed box. The display window is [0.01 0.030].

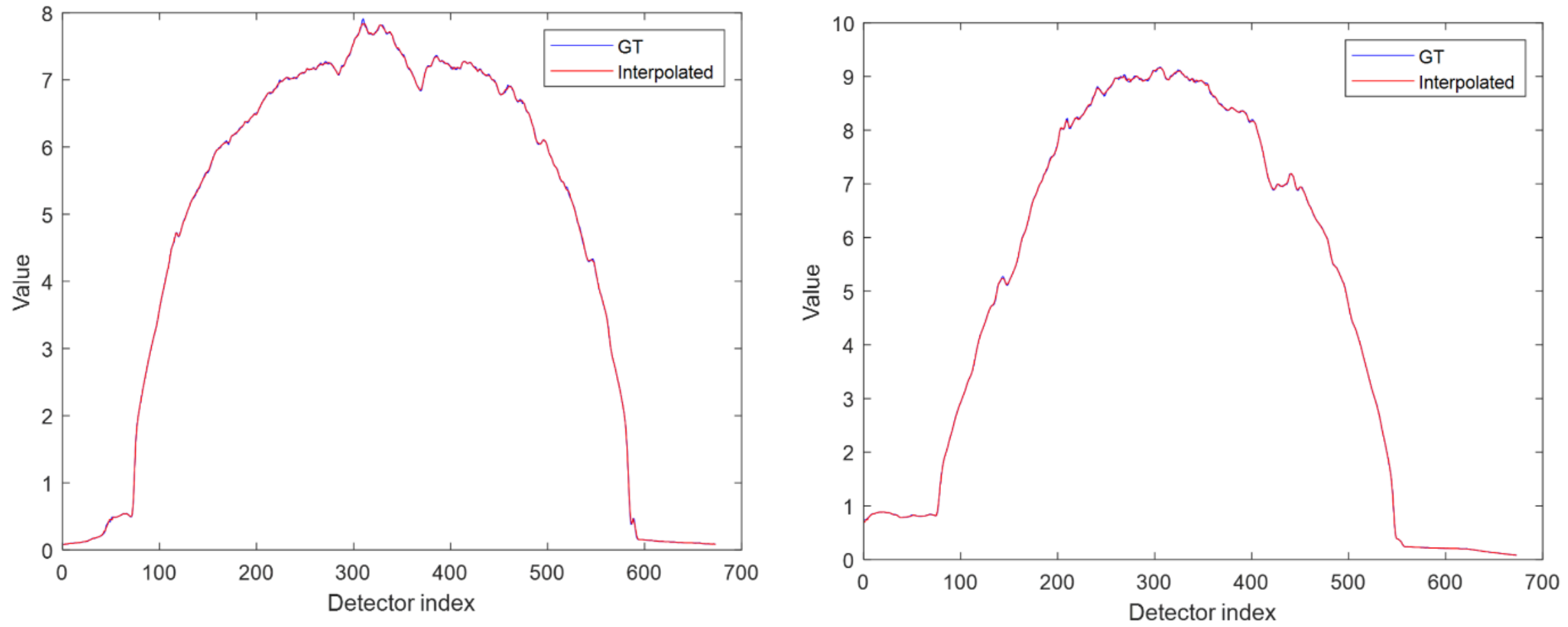

Fig. 15. 1/4 sparsity interpolation results of the Mayo data. From left to right, it showed results of 5th and 805th projections.

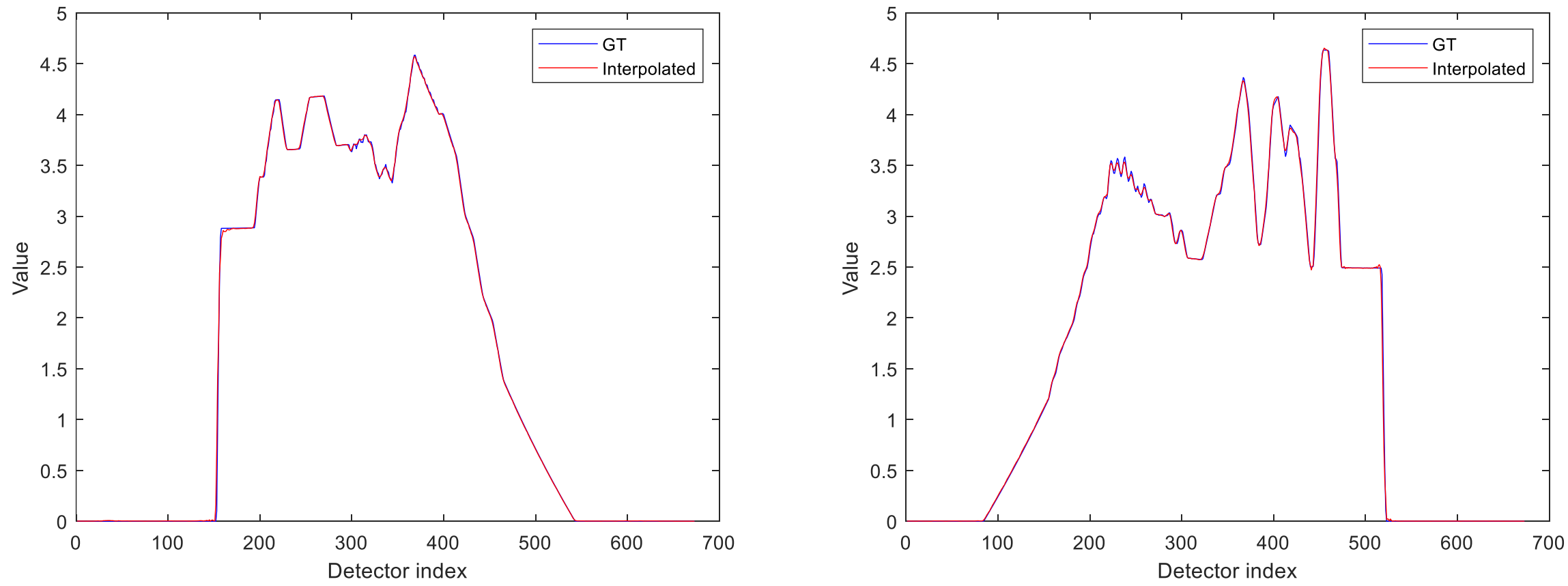

(a) (b)

Fig. 16. The interpolated results of resolution phantom for 1/8 sparse-view projection. Red line indicates the interpolated results and green line represented the original full view projection as ground truth (GT) (a). shows the results of 25th projection view. (b). shows the results of 121th projection view.

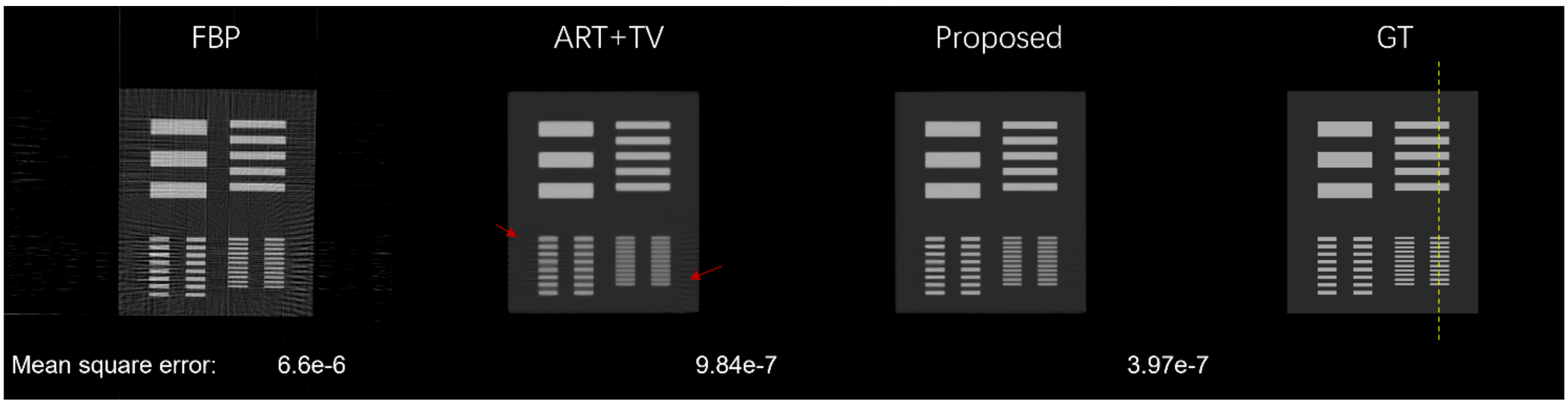

Fig. 17. The reconstructed results of the resolution phantom from 1/8 sparse-view projection. From left to right, it showed the reconstructed image of FBP, ART+TV, the proposed method, respectively. The display window is [0.05 0.035]. The red arrows indicated the obvious streak artefacts and the yellow dashed line represented the line profile location.

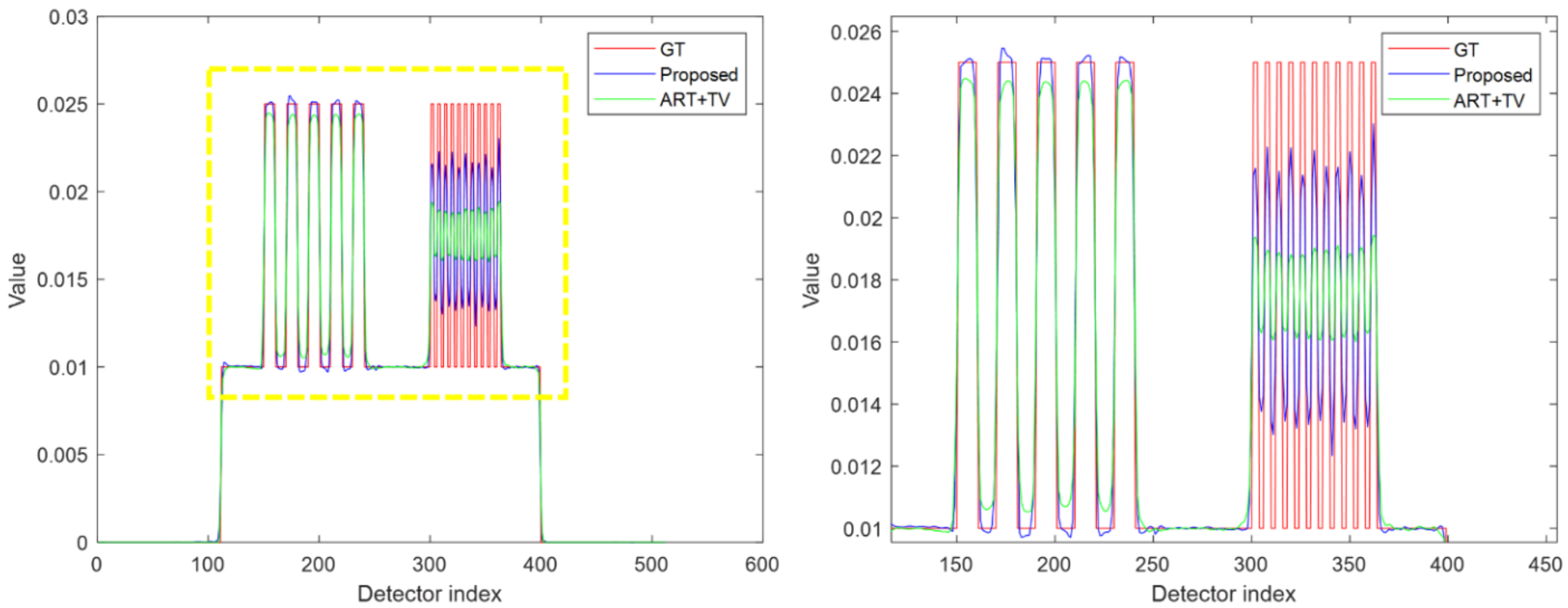

Fig. 18. The line profile of the yellow dashed line. Figure in right side is the enlarged area in yellow dashed box.

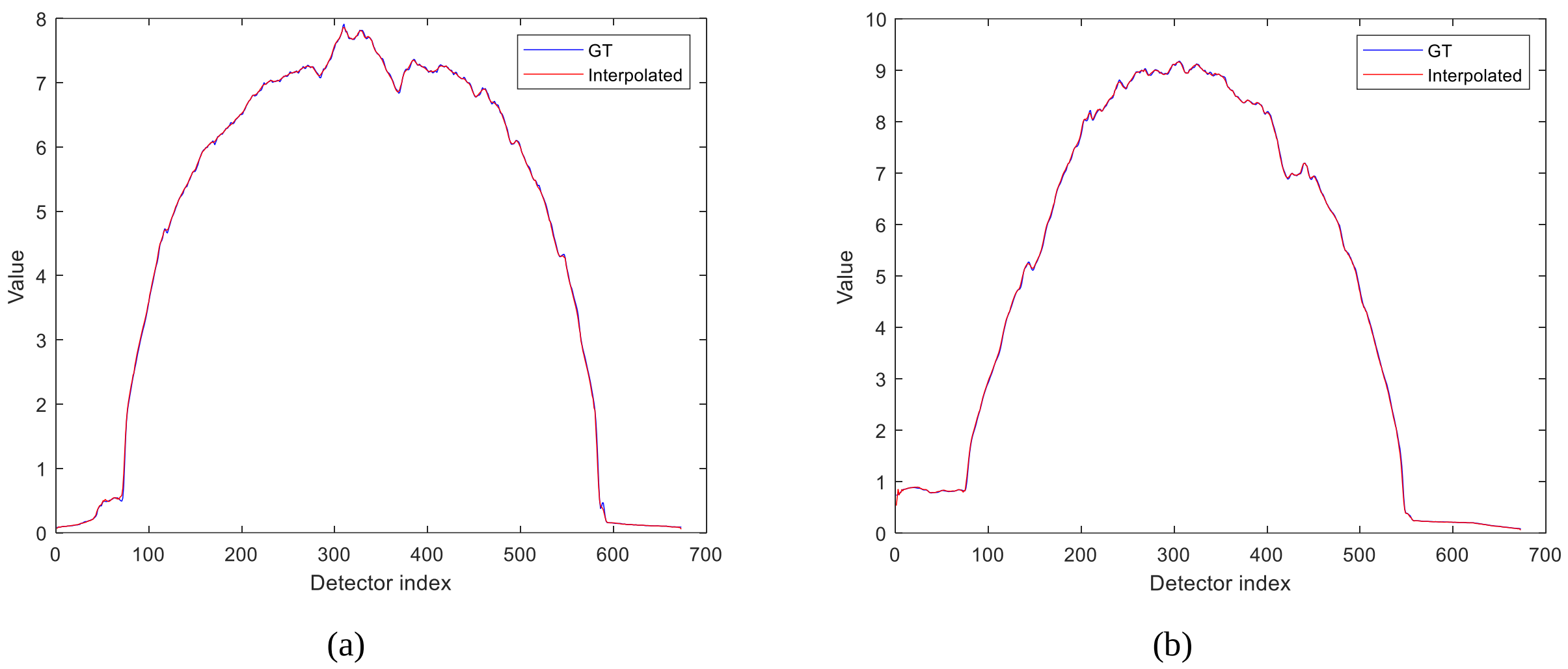

Fig. 19. The interpolated results of Mayo data of 1/8 sparse-view projection. Red line indicates the interpolated results and green line represented the original full view projection as ground truth (GT) (a). shows the results of 25th projection view. (b). shows the results of 121th projection view.

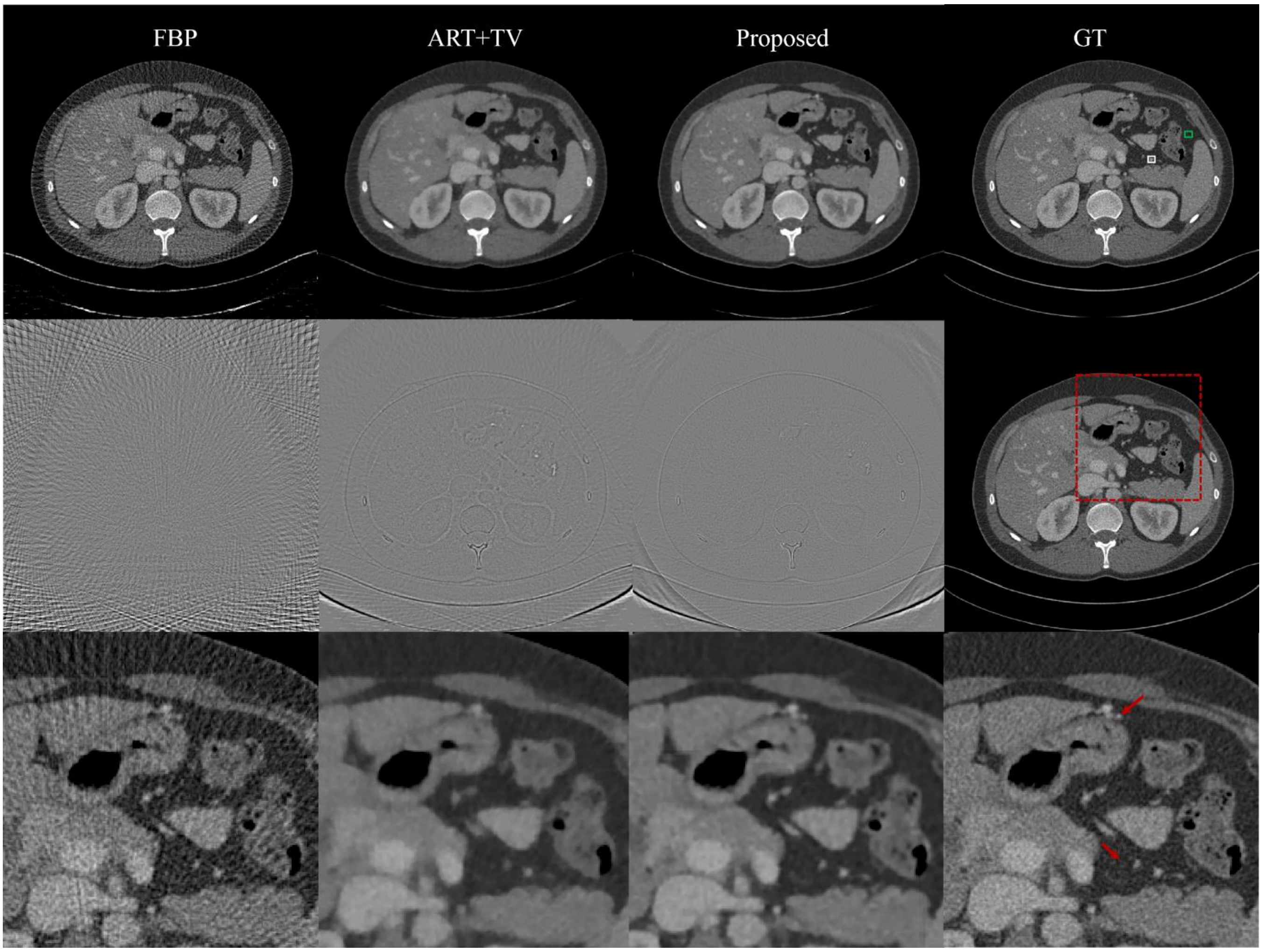

Fig. 20. The reconstructed results of a 1/8 sparse-view projection of Mayo data. First row showed the reconstructed image of FBP, ART+TV, the proposed method and GT, respectively. The display window is [0.01 0.030]. Second row described the residue image and selected zoomed in area. The display window of the residue image is [-1e-2 1e-2]. The third row is the enlarged area denoted in red dashed box.

## 5. Discussion and conclusion

In this study, we have derived a PDE of Radon transform, named LCE. The LCE is free of imaging object and reveals the universal property of Radon transform in terms of the correlation over a neighborhood of integral line. LCE is built on 2-D Radon transform domain. Specifically, LCE is a n-th order PDE of the Radon transform that involves in two angular derivates regarding to two rotation centers on its integral line, as well as derivate of the distance perpendicular to the line. For 3-D Radon transform cases, LCE can be applied in any 2-D profile of the 3-D Radon transform. In divergent beam CT, the LCE has its equivalent forms, i.e., cLCE and sLCE, when set one rotation center to the focal spot. In this case, the angular derivative regarding to the focal spot is easily obtained by the divergent beam projection, thus the cLCE and sLCE shown in Eqs.(17) and (23) can be directly used to model the correlation of projections among adjacent integral lines for actual CT systems.

Although it is a common sense that Radon transform contains redundancy, at present, there has not been a universal model proposed to describe the redundancy. The proposed LCE in this study can be considered as the first model for this purpose by disclosing the correlation of Radon transform over a neighborhood of integral line. On the basis of this regard, data completeness of CT projection in terms of enough scanning number could be no longer necessary since unmeasured data caused by under-sampling may be estimated by using the redundancy model. In fact, in this study, we have proposed a simple but efficient interpolation scheme to restore the unmeasured projection from a 1/4 sparse-view projection for the widely used circular scanning CT geometry, which only uses the first-order PDE for the interpolation. The conducted experiment of comparing the profiles of the restored projection with the ground truth, as well as the FBP reconstruction from the recovered projection have confirmed the proposed method. In this regard, we can expect that sparse-view scanning in CT contains enough information for precise reconstruction, which would bring extensive topics to be comprehensively investigated in future. In the following, we discuss some aspects that are related to the further work.

The proposed projection completeness method in this study is just based on the first order PDE and only use the angular derivates on the circular trajectory shown in Eq.(20). Indeed, projections along more intersected integral lines should be used for interpolation to utilize as abundant information as possible. Moreover, the second and higher order PDEs should also be used to achieve more precise interpolation results.

Though the LCE explains the redundancy existed in Radon transform, it is still a problem that how much sparse-view projection can be compensated by the LCE for a perfect reconstruction as from the complete projection. In this paper, we report a preliminary study that proposes a projection interpolation scheme using the discrete first-order LCE with the angular derivative on the circular trajectory and carries out a 1/4 sparse-view projection restoration experiment. In spite of satisfied recovering result achieved, this study just confirms that a sparse-view projection is possibly able to be completed by the LCE. However, a deep study should be taken to identify the lower bound of the sparse-view level up which any sparse-view projection can be completely compensated.

Implementing an algorithm must be based on discretization in practice, which unavoidably brings discretization errors. Since solving PDE is highly sensitive to errors, the proposed projection interpolation that directly uses the discretized form of the cLCE is inadvisable though it has achieved the satisfied interpolation effect in 1/4 sparse-view projection restoration. In this study, our purpose is to investigate whether sparse-view projection completeness is achievable based on the proposed LCE, hence the simple interpolation scheme is adopted. While in feasible applications, advanced signal process technologies, such as CS and deep networks, should be used to investigate reasonable solutions of applying the LCE.

The proposed LCE is built on the 2-D sinogram map. However, the LCE is also established on any 2-D profile of 3-D Radon transform. Hence, exploring a 3-D interpolation scheme for sparse-view cone beam scanning based on the LCE is also an attractive topic. Since the 3-D scheme can combine every 2-D profile of the 3D projection, more abundant information will be integrated. The early work to explore the correlation on cone beam projection are John's equation based schemes which have the limitation of trajectory condition, as summarized in section I. Compared with the John's equation based method, the LCE based scheme would not have the limitation and become a general interpolation scheme.

Nowadays, deep learning based technologies have been used for the projection completion (Lee *et al.*, 2018; Ding *et al.*, 2021). However, Radon transform is a dimensionality reduction function via line integral in especial. A natural 2-D or 3-D image including X-ray anatomic image, is consist of abundant of independent structure components that is easy to represent by a local feature operator or deep learning based network. While it is transformed by Radon transform, the line integration across various locally structural components that are independent to each other, makes the projection to appear no-structural property. Hence, Radon transform function loses the structural information by the dimensionality reduction of line integral. The non-structural image is hard to be represented by either local structure extractor or deep learning based network with high reliability and generalization ability. Nevertheless, the LCE and its equivalent forms provide structural representation of Radon transform, which could benefit to build a credible deep neural network with strong generalization ability to process the projection for either sparse-view interpolation or noise reduction.

CT image is sensitive to noise or error in projection because the reconstruction involves in a high pass ramp filtration. Meanwhile, solving the LCE is also sensitive to noise or error, even if advanced signal processing techniques can be applied to conduct a perfect interpolation scheme based on the LCE. Therefore, directly using the LCE to recover the sparse-view projection and following with a CT reconstruction based on the recovered projection is not a good idea, since noise and error are inevitably existed in the recovered projection. In this regard, combining the LCE based projection completeness and the CT reconstruction into a unified framework is a reasonable scheme. In the combination, the reconstruction part with regularization can help to suppress the error caused by the LCE based projection completeness, so that a unified framework would process higher sparse-view CT problem. Indeed, in this study, we propose such a unified reconstruction method shown in Figs.17-20. Although this method just using the first-order LCE and the angular derivates on the circular trajectory, as well as a simple TV regularization, the reconstruction shows appealing results which achieves 1/8 sparse-view CT reconstruction with satisfied image quality in the studied cases, demonstrating better performance compared to the two-step process of projection interpolation followed by FBP reconstruction that just attains 1/4 sparse-view CT reconstruction. It would be expected that by applying advanced signal processing technic and reconstruction method, such as deep learning based networks, the constructed unified reconstruction algorithm will realize more significant reconstruction results.

In conclusion, in this study, we propose the first universal n-th order PDE for Radon transform, named LCE, and deduce its equivalent forms for two typical CT geometry, i.e., cLCE for circular scanning trajectory and sLCE for stationary linear array scanning trajectory. The proposed LCE represents the redundancy property of Radon transform and suggests that sparse-view CT projection may contain enough information of complete projection to some extent. Hence, in practice the completeness requirement of CT scanning could be no longer needed. We also propose a simple but efficient projection interpolation method by using the discrete first-order LCE. Finally, we propose a unified algorithm for sparse-view CT reconstruction by combining the projection interpolation with regularized iterative reconstruction. The theoretical analysis, algorithm development, as well as the experiments in this study demonstrate an appealing topic in CT area: exploring the redundancy property of Radon transform and investigating new CT theories based on the redundancy property will boost the further development of CT reconstructions.

## Acknowledgments

We would like to thank Prof. Ge Wang with RPI and Prof. Hengyong Yu with UML for the helpful discussions on the study and their invaluable advice.